\documentclass[a4paper,aps,showpacs,showkeys,eqsecnum]{revtex4}

\usepackage{bm}
\usepackage{amsfonts}
\usepackage{amsmath}
\usepackage{amssymb}
\usepackage{graphicx}
\usepackage{feynmp}
\usepackage[usenames,dvipsnames]{xcolor}
\usepackage[colorlinks=true,linkcolor=Blue,citecolor=Blue,urlcolor=blue,linktocpage]{hyperref}

\newcommand{\ef}{w}

\newcommand{\aeq}{\begin{equation}}
\newcommand{\ceq}{\end{equation}}
\newcommand{\aec}{\begin{eqnarray}}
\newcommand{\cec}{\end{eqnarray}}
\newcommand{\ase}{\begin{subequations}}
\newcommand{\cse}{\end{subequations}}

\renewcommand{\(}{\left(}
\renewcommand{\)}{\right)}
\renewcommand{\[}{\left[}
\renewcommand{\]}{\right]}

\newcommand{\tensorrep}{\left(1,0\right)\oplus\left(0,1\right)}

\newcommand{\M}{\mathfrak{M}}

\renewcommand{\a}{\alpha}
\renewcommand{\b}{\beta}
\renewcommand{\c}{\chi}
\newcommand{\m}{\mu}
\newcommand{\n}{\nu}
\renewcommand{\o}{\omega}
\newcommand{\g}{\gamma}
\renewcommand{\d}{\delta}
\newcommand{\h}{\eta}
\newcommand{\z}{\zeta}
\newcommand{\f}{\phi}

\renewcommand{\l}{\lambda}
\newcommand{\s}{\sigma}
\renewcommand{\r}{\rho}
\newcommand{\q}{\theta}
\newcommand{\e}{\epsilon}
\newcommand{\x}{\xi}
\renewcommand{\t}{\tau}

\renewcommand{\k}{\kappa}

\newcommand{\G}{\Gamma}
\renewcommand{\P}{\Pi}

\newcommand{\pd}{\partial}

\newcommand{\gv}{g_V}

\begin{document}

\date{\today }

\title{Compton scattering off massive fundamental bosons of pure spin 1}

\author{E. G. Delgado-Acosta}\email{german@ifisica.uaslp.mx}
\affiliation{
Instituto de F\'{\i}sica, 
Universidad Aut\'onoma de San Luis Potos\'{\i},
Avenida Manuel Nava 6, 
San Luis Potos\'{\i}, 
San Luis Potos\'{\i} 78290, M\'exico}

\author{M. Kirchbach}\email{mariana@ifisica.uaslp.mx}
\affiliation{Instituto de F\'{\i}sica, 
Universidad Aut\'onoma de San Luis Potos\'{\i},
Avenida Manuel Nava 6, 
San Luis Potos\'{\i}, 
San Luis Potos\'{\i} 78290, M\'exico}

\author{M. Napsuciale}\email{mauro@fisica.ugto.mx}
\affiliation{Departamento de F\'{\i}sica, Universidad de Guanajuato,
Lomas del Campestre 103, Fraccionamiento Lomas del Campestre, Le\'on,
Guanajuato 37150, M\'exico}
\email{mauro@fisica.ugto.mx}

\author{S. Rodr\'{\i}guez}\email{simonrodriguez@uadec.edu.mx}
\affiliation{Facultad de Ciencias F\'{\i}sico Matem\'aticas, Universidad
Aut\'onoma de Coahuila,
Edificio "D," Unidad Camporredondo, Saltillo, Coahuila 25280, M\'exico}

\begin{abstract}
{
Relativistic particles with spins $J>0$ are described by means of multicomponent wave functions which transform covariantly according to Lorentz-group representations that contain at rest the spin of interest. The symmetry group of space-time provides not one but an infinity of such representations which are equivalent for free particles but yield different electromagnetic couplings upon gauging; thus the challenge is to  develop criteria  which allow us to select those of them which relate to physically detectable particles. We here take the position that the unitarity of the Compton scattering cross sections in the ultrarelativistic limit, when predicted by a consistent method for a spin-$1$ description, could provide such a criterion. We analyze the properties of massive fundamental spin-$1$ bosons transforming as antisymmetric tensors of second rank, $(1,0)\oplus(0,1)$. For this purpose, we employ the Poincar\'e covariant projector method, which provides consistent, gauge invariant, causal, and representation specific Lagrangians. This formalism  yields a twofold  extension of the Proca Lagrangian for the description of spin-$1$ bosons, first from an in-built $g=1$ value of the gyromagnetic ratio to an unspecified general $g\not=1$, and then from, single-parity, to parity-doublet degrees of freedom. We find different results for Compton scattering in these theories and track the differences to the lack of universality of the  vector-antisymmetric-tensor equivalence theorem which is specific only to Proca's framework, and valid for $g=1$,  while it is violated within the more general Poincar\'e covariant projector formalism. Our main result is  that a finite Compton scattering differential cross section in the ultrarelativistic limit requires us to consider the contributions of both parities in $(1,0)\oplus(0,1)$. On that basis, we conclude that massive spin-$1$ bosons transforming as antisymmetric tensors are physical parity doublets.
}
\end{abstract}

\maketitle

\tableofcontents
\section{\label{sec1}Introduction}
The description of particles with higher  spins, $J>1/2$, continues to be among the challenges in contemporary theoretical physics. According to the standard strategy \cite{Weinberg:1995mt,Weinberg:1996kr,Weinberg:2000cr} such particles are described by means of multicomponent wave functions which transform covariantly according to Lorentz-group representations that contain at rest the spin of interest. The symmetry group of the Minkowski space-time provides infinitely many nonequivalent representations of the required property and it is necessary to develop criteria for the selection of those few representations which describe physically detectable particles. The unitarity of the Compton scattering amplitudes in the ultrarelativistic limit could serve as such a criterion.  It is the goal of the present work to study the question on the observability of massive fundamental bosons of spin-$1$ transforming according to the Lorentz-group representation, $(1,0)\oplus(0,1)$. 
\noindent
The pure spin-$1$ representation space of the Lorentz group is among the most important covariant entities in physics. In any gauge theory, be it Abelian as the electrodynamics, or, non-Abelian as QCD and the electroweak theory, it describes the antisymmetric field tensor of second rank, in terms of which the respective Lagrangians are written. Also the six Lorentz group generators in any arbitrary representation have the property to transform as the components of an antisymmetric tensor of second rank. Moreover, the massless representation under discussion can be employed in the description of gauge fields between extended objects, such as strings \cite{Kalb:2009fj}. Specifically according to \cite{Kalb:2009fj}, and references therein, the spin-$1$ massless particles described by means of the antisymmetric tensor carry only 1 degree of freedom of zero helicity, complementary to the massless vector gauge field, characterized by 2 transverse degrees of freedom. These so-called Kalb-Ramon gauge fields are highly significant in the theory of open strings,  where they underlie the important Green-Schwarz anomaly cancellation mechanism. Through their interactions with  ordinary gauge fields, associated with the four-vectors between the open ends of the strings, a tensor-vector coupling can be designed  which can trigger a mass generation \cite{Kalb:2009fj} alternative to the Higgs mechanism, which is based on scalar-vector couplings. 

Switching to massive spin-$1$ particles does not bring physics of  the four-vector and antisymmetric-tensor particles closer; they still remain apart in several aspects. It is the goal of the present study to reveal some of  those aspects. 

That spin-$1$ antisymmetric-tensor mesons are quite distinct from the four-vector mesons, has been demonstrated in detail for the case of composite hadrons in Ref.~\cite{Chizhov_review} with the emphasis on their  weak decay, and strong  couplings. There, new relationships between the squared masses of $1^{--}$ vector mesons ($\rho$'s), on the one side, and axial-vector  $1^{+-}$ and $1^{++}$ mesons, ($b_1$ and $a_1$, respectively) on the other, have been found. We here instead focus on pointlike massive spin-$1$ bosons, with the emphasis on their electromagnetic properties.  We recall that massive $(1/2,1/2)$ particles are endowed with 4 degrees of freedom distributed over one scalar and three vectors of opposite parities and opposite signs of the norms, making the scalar degree of freedom into a redundant component. Instead, the massive antisymmetric tensor carries 6 well behaved degrees of freedom distributed over a pair of spin-$1$ bosons constituted by a  polar and an axial vector,  equivalently, over a parity doublet. It finds  applications ranging from  effective theories \cite{Ecker:1989yg} over radiative pion decay to cosmology \cite{Chizhov_review}. For all these reasons, expanding knowledge on the properties of particles of pure spin $1$ appears timely.

The present investigation is devoted to the study of Compton scattering off a massive antisymmetric-tensor field in comparison with the same process involving as a target a massive four-vector field like the electroweak $W$ gauge boson. We use  Lagrangians designed within the framework of the recently elaborated \cite{Napsuciale:2006wr} Poincar\'e  covariant projector method where the spin-$1$  Lagrangians contain the generators of the representation under investigation and therefore, in general, allow for detecting differences in  the properties of particles transforming by different Lorentz boosts. These Lagrangians are moreover consistent insofar as they lead to hyperbolic wave equations which describe causal propagation of the wave fronts in the presence of an electromagnetic field. They can be viewed as extensions of Proca's Lagrangians in the four-vector and the antisymmetric-tensor representations from the in-built unphysical $g=1$ value for the gyromagnetic ratio toward a general $g\not=1$ which is specified to its physical value upon introducing interactions and from some suitably chosen dynamical constraints. Moreover, the spin-$1$ Lagrangian for the antisymmetric-tensor representation arising within the Poincar\'e covariant projector method describes a parity doublet, conversely to the Proca Lagrangian, which confines us to the degrees of freedom of only one parity. These extensions are important insofar as the  $(1/2,1/2)$-$(1,0)\oplus(0,1)$ indistinguishability arising within Proca's theory and known under the name of vector-tensor equivalence theorem has its validity  only for $g=1$,  and is, in general, removed within the Poincar\'e covariant projector method allowing for $g\not=1$. In this fashion, the four-vector and the antisymmetric-tensor representations become distinguishable and can be populated by fundamental bosons of distinct physical properties. Which specifically, will become clear in due course.\\
In particular, in a work prior to this \cite{DelgadoAcosta:2012yc} we reported on different electromagnetic properties of massive four-vector and antisymmetric-tensor bosons. Namely, we calculated the electric quadrupole  moments of such particles and found them  distinct by a sign, assuming equal values of the gyromagnetic ratio, $g$. The difference in question  has been found to be of pure kinematic origin and has been attributed to the different orientations of the electric quadrupole moments in the Breit frame, oblate for $(1/2,1/2)$, versus prolate  for $(1,0)\oplus (0,1)$. In the present work we are studying the impact of the above difference on the Compton scattering amplitudes off fundamental $(1,0)\oplus (0,1)$ bosons. Our main result is that finite  differential and total cross sections in the ultraviolet are obtained only for scattering off the parity doublet as a whole.

The paper is organized as follows. The first part represents a detailed comparison between two formalisms for a spin-$1$ description, namely, Proca's approach and the Poincar\'e covariant projector method. The aim is to reveal the credibility and physical predictive power of the latter method and motivate its subsequent employment in the analysis of the observability of fundamental single-spin-$1$ bosons. The second part is entirely devoted to the calculation of Compton scattering off $(1,0)\oplus (0,1)$. Specifically, in the next section we present four spin-$1$ wave equations,  all  equivalent in the absence of interactions though  well distinguishable through the related Noether currents and  upon gauging. Two of them are the standard Proca equations for $(1/2,1/2)$ and $(1,0)\oplus(0,1)$, respectively, which we here rederive from combined mass-$m$ and parity projectors, while the other two are their counterparts following from the Poincar\'e covariant projector method \cite{Napsuciale:2006wr}. We analyze hyperbolicity and causality of the former wave equations upon gauging showing that within $(1,0)\oplus (0,1)$ the negative parities are described by a fourth order differential  equation whose hyperbolicity is inconclusive from the perspective of the Gilbert-Currant criterion, while the positive parities are described by a hyperbolic and causal second order equation.

This is at variance with the case of the spin-$1$ description in terms of the four-vector $(1/2,1/2)$, where one finds the negative-parity states satisfying a hyperbolic genuine gauged equation, while the attempt to find a genuine gauged equation for the scalar sector leads to a fourth order equation and of an inconclusive hyperbolicity within the context of the aforementioned Gilbert-Currant criterion. Compared to this, the  $(1,0)\oplus(0,1)$ wave equations following  from the Poincar\'e covariant projector method are manifestly and unconditionally hyperbolic and causal  for all involved degrees of freedom. In the same  section  we furthermore discuss the vector-tensor equivalence theorem and its violation within the Poincar\'e covariant projector method at both the levels of the electromagnetic currents and the genuine gauged wave equations. Section \ref{c1001} is devoted to the calculation of the Compton scattering off the single-parity degrees of freedom in the antisymmetric-tensor representation, showing a differential cross section that grows infinitely with the energy increase, except in forward direction. In the same section we also present the calculation of  Compton scattering off the $(1,0)\oplus (0,1)$ parity doublet as a whole which allows us to obtain finite differential and total cross sections in all directions. The paper closes with brief conclusions.


\section{Wave equations for spin 1}
Sometimes one may be prompted to  believe that the properties of a particle with spin $J$ are independent of the Lorenz-group representation employed in its description. Take for concreteness the case of spin 1.  One well-established correct statement in the literature is the so-called  vector-tensor equivalence theorem, whose principal meaning is  that the construction of the solutions of Proca's equation for a noninteracting spin-1 boson residing  in $(1,0)\oplus (0,1)$ reduces to the construction of the solutions of a boson of equal spin described as a four-vector. As we shall demonstrate below, this equivalence extends to the genuine gauged equations and the electromagnetic multipole moments of their currents, as well. Yet, as it will become clear in due course, this is not a universal property.

In order to gain a deeper insight in this issue, we here recall the standard procedure for the construction of equations of motion for particles residing within a given Lorentz-group representation, spanned by the generic degrees of freedom, $\psi({\mathbf p}, \lambda)$, where ${\mathbf p}$ is the three momentum, and $\lambda$  denotes  the set of quantum numbers characterizing the degrees of freedom under consideration. A wave equation within a given Lorentz-group representation (ordinarily one is dealing with finite-dimensional representations) is straightforwardly obtained from a projector, call it $\Pi_{\lbrace \kappa  \rbrace} ({\mathbf p})$, where $\lbrace \kappa  \rbrace $ is a generic representation labeling, on the basis vectors, $\psi ({\mathbf p},\lambda)$ in the representation space of interest, according to
\begin{equation}
\Pi_{\lbrace \kappa \rbrace} ({\mathbf p})\psi({\mathbf p},\lambda )
=\psi ({\mathbf p},\lambda ).
\label{waqu_Prjt}
\end{equation}
For example, the free Klein-Gordon equation is obtained exclusively in  terms of the  mass-$m$ projector as
\begin{equation}
\Pi_{0}({\mathbf p})\phi ({\mathbf p})=\frac{p^2}{m^2}\phi ({\mathbf p})=\phi
({\mathbf p}),
\end{equation}
while the celebrated Dirac equation for the electron follows  from exclusively projecting on the positive parity degrees of freedom in the $\kappa:\,\, (1/2,0)\oplus (0,1/2)$ representation,
\begin{equation}
\Pi_{(1/2,0)\oplus(0,1/2) } ({\mathbf p})u ({\mathbf p},\lambda )=u({\mathbf
p},\lambda ), \quad
\Pi_{(1/2,0)\oplus (0,1/2) } ({\mathbf p})=\frac{p\!\!\!/ +m}{2m}, \quad
\lambda=\pm \frac{1}{2},
\label{waDrc}
\end{equation}
where $\Pi_{(1/2,0)\oplus (0,1/2) } ({\mathbf p})$ is the covariant positive-parity projector. As we shall see below, the projector $\Pi_{\lbrace \kappa\rbrace }({\mathbf
p})$ is quite general and can also be  a product of mass-$m$, parity, spin, etc.  projectors.
 
\noindent
For the case of spin 1 of interest here, two Lorentz group representations are of basic  use. These are the four-vector, $(1/2,1/2)$, on the one side,  and the
antisymmetric tensor, $(1,0)\oplus (0,1)$, on the other. As long as the choice of the projector in (\ref{waqu_Prjt}) is not unique, different wave equations, all equivalent at the noninteracting level,  can be constructed.

In the subsequent section  we address the following three problems regarding the description of physical  spin-1 bosons:
\begin{itemize}
\item the nonuniqueness of the free-particle wave equations within the same representation, be it the four-vector or the antisymmetric tensor, and its consequences for the electromagnetic properties of the particles,

\item the general representation dependence of the wave equations and the associated electromagnetic  currents,

\item the so-called vector-tensor  equivalence theorem  for spin 1 \cite{Jenkins:1972pd} and its lack of universality.
\end{itemize}

\subsection{Nonuniqueness of the free-particle wave equations for spin 1 as a four-vector}

The four-vector Lorentz group representation, $(1/2,1/2)$, usually considered to be of natural parity, has been extensively studied and it is a well-known fact that it is spanned by four basis states, one of which is a timelike scalar, while the remaining  three are spacelike spin-$1$ polar vectors. The four-vector therefore contains two sectors of different spins and opposite parities. If one wishes to employ it in the description of spin-1 bosons,  the scalar sector has to be projected out. This is not a  difficult task because the vector degrees of freedom happen to be simultaneously those of negative parities and as shown below, they can  easily be identified  by means of a corresponding covariant parity projector.

\subsubsection{Deriving Proca's equation from a combined mass-\texorpdfstring{$m$}\space{},  and negative-parity projector: Causality proof}
A spin-$1$ wave equation for a particle transforming in the four-vector representation, $(1/2,1/2)$, of the Lorentz group, can be constructed along the line of \eqref{waqu_Prjt} considering $\Pi_{(1/2,1/2)}({\mathbf p})$ as a combined mass-$m$ and negative-parity projector according to
\begin{equation}
\Pi_{(1/2,1/2)}^{\a\b}({\mathbf p})=\frac{{\widehat P}^2}{m^2}{\mathbf P}_{-}^{\alpha\beta}({\mathbf p}),
\label{fourvprj}
\end{equation}
where ${\widehat P}$ denotes the operator of four-momentum,  $m$ stands for the particle mass, ${\widehat P}^2/m^2$ is the mass-projector operator, and ${\mathbf P}_{-}^{\alpha \beta}({\mathbf p})$ is the covariant projector on the negative-parity states. The latter is well known and is straightforwardly constructed as
\aeq
\mathbf{P}_-^{\a\b}({\mathbf p})=\sum_{\l}\eta^{\alpha}({\mathbf p},\l)[\eta^{\beta}({\mathbf p},\l)]^* =g^{\a\b}-\frac{p^\a p^\b}{p^2},
\qquad \l=-1,0-1,
\ceq
where $\eta^{\alpha}({\mathbf p},\l)$ denote the boosted negative-parity basis
states in  $(1/2,1/2)$
(see \cite{Napsuciale:2006wr} and \cite{DelgadoAcosta:2012yc}, among others, for details). The action of the and mass-shell projector operators on the states under consideration is given as follows:
\aec
\mathbf{P}_{-}^{\a\b}({\mathbf p})\eta_\b({\mathbf p},\l )=\eta^\a({\mathbf p},\l), &\quad& 
\frac{{\widehat P}^2}{m^2} \eta_\a ({\mathbf p},\l ) =\eta_\a ({\mathbf p},\l ).
\cec
With that, and introducing the quantized field, $V^\a(x,\l )= \int {\mathrm d}^4p \left(\h^\a({\mathbf p},\lambda )e^{-ix\cdot p}a _\lambda ({\mathbf p})
+ [\h^{\a}({\mathbf p},\lambda)]^*a^\dagger_\lambda ({\mathbf p}) e^{i x\cdotp}\right)$, according to \cite{Bjorken}, Eq.~(\ref{fourvprj}) takes the following familiar form in  position space:
\aec
(\pd^2+m^2)V_\a (x,\l)-\pd_\b \pd_\a V^\b(x,\l)&=&0\label{parpro_proca}.
\cec
As a reminder, there are three  spin-$1$ four-vectors, $V_\alpha (x,\l)$, distinct by the $\l$ values. Now it is quite instructive to compare the latter equation to the standard one following from Proca's Lagrangian \cite{APro}, in which the $\l$ label is usually suppressed (we also will suppress the position argument for the sake of simplifying notations),
\begin{equation}\label{Proca}
{\mathcal L}_P=-\frac{1}{4}[U^{\dagger}]^{\alpha\beta }U_{\alpha \beta}+\frac{1}{2}m^2 [V^{\dagger}]^\alpha V_\alpha,\\
\end{equation}
with $U_{\alpha \beta}=\partial _\alpha V_\beta -\partial _\beta V_\alpha $ and $\alpha=0,1,2,3$. The resulting Euler-Lagrange equation is then,
\aec\label{prowafu}
\left( \partial ^2+m^2\right)V_\alpha  -\pd_\b \partial_\a V^\b&=&0.
\cec
A Comparison of (\ref{prowafu}) and (\ref{parpro_proca}) shows that Proca's formalism for spin 1 is no more than the projection on negative-parity basis states of $(1/2,1/2)$ of mass $m$. The same scheme can be applied to the scalar state of positive parity, here denoted by $V^4_\alpha $,
in which case one encounters a nonpropagating state according to
\aec
\pd_\b \pd_\a [V^4]^\b&=&-m^2V^4_\alpha.\label{nonprop_1}
\cec
Caution is necessary in the interpretation of the gauged equations, (\ref{parpro_proca}) and (\ref{nonprop_1}). Indeed, upon naive gauging, i.e. upon simply replacing the ordinary derivatives, $\partial_\mu$, by the covariant ones, $D_\mu=\partial_\mu +ieA_\mu$, specifically the Proca equation, (\ref{parpro_proca}) becomes
\begin{equation}
\left[g^{\alpha}{}_{\beta}(D^2+m^2) -D^\alpha D_\beta
+ieF^{\alpha}{}_{\beta}\right]V^\beta =0.
\label{Proca_ggd}
\end{equation}
This is a $4\times 4$ matrix equation with the (derivative) "coefficients" of $V^\beta$ arranging to  a matrix, call it ${\mathbf A}^\alpha\, _\beta (D)$, and given by
\begin{equation}
{\mathbf A}^\alpha{}_\beta (D)=g^\a{}_{\b} D^2-D^\alpha D_\beta.
\label{Coeff_mtrx}
\end{equation}
It is easy to see that the equation (\ref{Proca_ggd})  is not a genuine one because it does not contain the time derivatives of $V^0$; this because of the following cancellation in the ${\mathbf A}^0{}_0\left( D\right)$ element:
\begin{eqnarray}
{\mathbf A}^0\, _0\left( D\right) &=& g^0\, _0\left( \left( D^0\right)^2
-\left(D^1\right)^2 -\left( D^2\right)^2 -\left( D^3\right)^2 \right)
-D^0D_0\nonumber\\
&=&-\left(D^1\right)^2 -\left( D^2\right)^2 -\left( D^3\right)^2.
\end{eqnarray}
A remedy to this shortcoming of (\ref{Proca_ggd}) is provided \cite{Velo:1970ur} by reincorporating in it the gauged auxiliary condition,
\begin{equation}
-m^2D\cdot V=\left(D_\beta D^2 - D_\mu D_\beta D^\mu\right)V^\beta.
\end{equation}
Doing so amounts to the following genuine differential equation:
\begin{equation}
\left[g^\alpha{}_\beta \left( D^2 +m^2\right) -D^\alpha \frac{ie}{m^2}F_{\mu\beta}D^\mu +i e F^{\alpha}{}_{\beta}\right]V^\beta =0,
\label{gut_gg}
\end{equation}
where use has been made of $\left[ D_\mu, D_\nu\right]=ieF_{\mu\nu}$. It is now quite instructive to check whether the solutions of the latter equation describe causally propagating waves, in which case it  would be hyperbolic. The hyperbolicity of differential and maximally second order equations  is tested  by means of the well-known Currant-Gilbert criterion which prescribes the calculation of the so-called characteristic determinant. The latter  is no more than $|{\mathbf A}\left (D\right)|$
but with the derivatives  being replaced by some constant four-vectors, here denoted by   $N^\alpha$, and $N_\beta$, according to,  $D^\alpha D_\beta  \to N^\alpha  N_\beta  $. The roots, $n_\mu$, of this determinant,  i.e. $|{\mathbf A}\left(n\right)|=0$, can then  be given the interpretation of the normals to the characteristic cones.
Then the  equation is hyperbolic if the timelike components, $n_0$, of all the normal vectors happen to be  real.
Applying the Currant-Hilbert criterion to (\ref{Proca_ggd}) one finds
\begin{equation}
|{\mathbf A}\left(n \right)|=n^6\left( n^2 -\frac{ie}{m^2}n^\mu
F_{\mu\nu}n^\nu\right)=\left(n^2\right)^4 =0.
\label{Proca_chdt}
\end{equation}
It is obvious that the determinant nullifies for real timelike components of the vectors normal to the characteristic cones, $n_0=\pm \sqrt{n_1^2+n_2^2+n_3^2}$, thus proving the hyperbolicity of the the gauged Proca equation. This is in reality a well-known fact which we presented here as a preparative example. Below we shall follow the same scheme to test the less studied hyperbolicity of the Proca equations in $(1,0)\oplus (0,1)$ and draw some nontrivial conclusions.

Instead, the genuine gauged equation corresponding to (\ref{nonprop_1}) for the scalar sector, turns to be of fourth order,
\begin{equation}
\left[ g_{\a\b}m^2 -\frac{1}{m^2}D_\a (-ie D^\s F_{\s\b}+D^2 D_\b)-ie
F_{\a\b}\right]V^\b _+=0,
\label{bad_gg}
\end{equation}
and its hyperbolicity is inconclusive from the perspective of the Gilbert-Currant criterion.

\subsubsection{Spin-1 wave equation from the Poincar\'e covariant projector method: Causality proof}
A further equation can be designed on the basis of a Poincar\'e covariant spin-1 and mass-$m$ projector,
\begin{equation}
\Pi^{\alpha\beta}_{(1/2,1/2)}({\mathbf p})=-\frac{1}{2m^2}\left(W^2_{(1/2,1/2)}({\mathbf p})\right)^{\alpha\beta},
\label{Prime_one}
\end{equation}
where $W^2_{(1/2,1/2)}$ is the squared Pauli-Lubanski vector in $(1/2,1/2)$, whose action on the spin-$1$ sector in the four-vector is given by
\begin{equation}
\left( W^2_{(1/2,1/2)}\right)_{\alpha \beta}\eta^\beta  ({\mathbf p},\lambda)=-2m^2\eta_\alpha({\mathbf p},\lambda ).
\label{PL_spin1}
\end{equation}
The explicit expression for the Pauli-Lubanski vector in the representation space under investigation is
\begin{equation}
\left( W_{(1/2,1/2)}\right)_\lambda =\frac{1}{2}\epsilon_{\lambda
\rho\sigma\mu}M_{(1/2,1/2)}^{\rho\sigma}p^\mu,
\quad \left[ M_{(1/2,1/2)}^{\rho\sigma}\right]_{\alpha\beta}=i(g^\rho{}_\alpha
g^\sigma{}_\beta -g^\rho
{}_\beta g^\sigma{}_\alpha),
\label{PL_FV}
\end{equation}
where $M_{(1/2,1/2)}^{\rho\sigma}$ are the generators of the Lorentz group in the four-vector. The squared Pauli-Lubanski vector is then calculated as
\begin{eqnarray}
\left( W^2_{(1/2,1/2)}\right)_{\alpha\beta}=-2(g_{\alpha\beta }g_{\mu\nu}-g_{\alpha\nu}g_{\beta\mu})p^\mu p^\nu &=&-2(g_{\alpha\beta}p^2 -p_\beta
p_\alpha).
\label{PLFV_SQRD}
\end{eqnarray}
Substitution of the latter equation in (\ref{Prime_one}), and taking into account (\ref{PL_spin1}), results in
\begin{equation}
(p^2-m^2)\eta_\alpha ({\mathbf p},\lambda) -p_\beta p_\alpha \eta^\beta({\mathbf p},\lambda ) =0,
\label{step_1}
\end{equation}
which is nothing more than the momentum space form of (\ref{parpro_proca}). This is not astonishing given the fact that the negative-parity basis spanning $(1/2,1/2)$ is vectorial; i.e. they are spin-$1$ states at rest, meaning that the squared Pauli-Lubanski operator tracks down the same degrees of freedom in the four-vector as the negative-parity projector. Yet, compared to (\ref{parpro_proca}), a new element can be  added to (\ref{step_1})  by the replacement,
\begin{equation}
p_\beta p_\alpha=\frac{1}{2}\lbrace p_\alpha ,p_\beta\rbrace-\frac{1}{2}\lbrack p_\alpha, p_\beta\rbrack,
\end{equation}
in which case Eq.~(\ref{step_1}) becomes
 \begin{equation}
(p^2-m^2)\eta_\alpha({\mathbf p},\lambda ) -\frac{1}{2}\lbrace p_\alpha,p_\beta\rbrace \eta^\beta({\mathbf p},\lambda )
+\frac{1}{2}\lbrack p_\alpha, p_\beta\rbrack \eta^\beta ({\mathbf p},\lambda ) =0.
\label{step_2}
\end{equation}
In the latter equation, an essential new ingredient has been built in through the commutator between the two momenta involved, which is vanishing for free particles, but contributes the electromagnetic field tensor upon gauging according to
\begin{equation}
\left[ \pi^\alpha , \pi^\beta\right]=-i\,e\,F^{\alpha \beta}, 
\quad\pi^\mu = p^\mu -\,e\,A^\mu.
\label{gauging}
\end{equation}
Here,  $A^\mu$ stands for the electromagnetic four-vector potential, and $e$ denotes the electric charge. A last step toward the Poincar\'e covariant projector method is extending the antisymmetric part of the equation by the most general antisymmetric tensor in $(1/2,1/2)$, which is $M_{\alpha\beta} $  from (\ref{PL_FV}), weighted by an arbitrary parameter, here denoted by $g_V$. Making use of $(-i)[M^{\mu\nu}_{(1/2,1/2)}]_{\a\b}p_\mu p_\nu=\left[p_\a,p_\b\right]$ allows us to extend the Eq. (\ref{step_2}) toward
\begin{equation}
(p^2-m^2)\eta_\alpha({\mathbf p},\lambda ) -\frac{1}{2}\lbrace p_\alpha
,p_\beta\rbrace  \eta^\beta ({\mathbf p},\lambda )+
\left(g_V-\frac{1}{2}\right)\lbrack p_\alpha , p_\beta\rbrack \eta^\beta
({\mathbf p},\lambda )=0.
\label{procaext}
\end{equation}
Equation (\ref{procaext}) appears as the Euler-Lagrange equation of the following  Lagrangian in position space:
\begin{eqnarray}
{\mathcal L}^V&=&-(\partial^\mu V^\alpha )^{\dagger}\Gamma^V_{\alpha
\beta\mu\nu}\partial ^\nu
V^\beta +m^2 V^{\alpha \dagger}V_\alpha , \label{PCPF}\nonumber\\
\Gamma^V_{\alpha\beta}{}^{\mu\nu}&=&g_{\alpha \beta }g^{\mu\nu}
-\frac{1}{2}\(g_{\alpha}{}^{\nu}g_{\beta}{}^{\mu}+g_{\alpha}{}^{\mu}g_{\beta}{}^{
\nu}\)
-i\({ g_V}-\frac{1}{2}\)[M^{\mu\nu}_{(1/2,1/2)}]_{\alpha\beta}\label{V_tnsr},
\end{eqnarray}
(see \cite{Napsuciale:2006wr} for more technical details). Before closing this subsection it is quite convenient  to obtain the gauged equation (\ref{procaext}), along the line of Eqs.~(\ref{Proca_ggd})-(\ref{gut_gg}). This amounts to
\begin{equation}
\left[
 g^\alpha \, _\beta \left( D^2 +m^2\right) +
\frac{ie}{m^2}D^\alpha \left(
(g_V-2) F_{\mu\beta}D^\mu + (g_V-1)\partial ^\mu F_{\mu\beta}\right)
+ieg_V F^{\alpha}\, _{\beta}\right]V^\beta =0.
\label{gut_gg_NKR}
\end{equation}
Though Eqs.  (\ref{gut_gg_NKR}) and (\ref{Proca_chdt}) are not equivalent, their characteristic determinants, in being defined only by the highest order derivatives, are,  despite the appearance of the  more general weight, $(g_V-2)ie/m^2$,  of the identically vanishing $ n^\mu F_{\mu\nu}n^\nu $-term. In other words, 
the hyperbolicity of the equation of motion is in the first instance, guaranteed by the antisymmetric nature of $F_{\mu\nu}$ and does not necessarily require $g_V=2$, though the latter value would have same effect the determinant.

\noindent
Notice that

\begin{itemize}

\item the hyperbolicity condition is valid for any value of the gyromagnetic ratio, $g_V$, and, differently from the case of spin $3/2$ in the four-vector spinor, earlier considered in \cite{Napsuciale:2006wr}, ceases to restrict it,

\item the explicit dependence of the Lagrangian in (\ref{PCPF}) on the Lorentz-algebra generators makes it representation specific and in this way suited for studying the
differences between  the properties of particles transforming by different boosts, a subject studied in Sec \ref{s2b} below.

\end{itemize}

However, for the time being we keep staying in the $(1/2,1/2)$ representation, and first attend to the distinction between the electromagnetic currents in Proca's approach and the Poincar\'e covariant projector method.

\subsubsection{Distinction of the spin-1 Lagrangians with \texorpdfstring{$g_V=1$}\space{} and 
\texorpdfstring{$g_V\not=1$}\space{} through their respective electromagnetic currents}

It is obvious from the free particle equations, (\ref{prowafu}) and (\ref{procaext}), that Proca's  theory and the Poincar\'e covariant projector method become equivalent for $g_V=1$. However, as we shall demonstrate below, for $g_V\not=1$ and upon gauging this equivalence is removed  and the two equations under discussion become profoundly different. On the one side, such differences will show up in the respective solutions, left aside in the present work, and on the other side they will become evident through the multipole decompositions of the related Noether currents, a subject treated in this subsection.  To be specific, the electromagnetic  current  corresponding to the Poincar\'e covariant projector method, here given in momentum space, is  obtained as
\aeq
j^V_{\m}(\mathbf{p}',\lambda^\prime;\mathbf{p}, \lambda)
=-e \, [{\h^{\a}}(\mathbf{p}',
\lambda')]^*(\G^V_{\a\b\n\m}p'^\n+\G^V_{\a\b\m\n}p^\n)
\h^\b(\mathbf{p}, \lambda ).
\label{curr1}
\ceq
Because   $p^\a\h_\a(\mathbf{p},\lambda )=0$ holds on the mass shell, and inserting  the explicit form of $\G^V_{\alpha\beta\mu \nu}$ in (\ref{V_tnsr}), the expression for the current simplifies to 
\begin{eqnarray}\label{gordonv}
\left( j^V\right) ^ {\m}(\mathbf{p}',\lambda^\prime;\mathbf{p},  \lambda )
&=&-e \, [\h^{\a}(\mathbf{p}',\lambda^\prime)]^*
[P^\m g_{\a\b}+i\gv [M_{(1/2,1/2)}^{\m\n}]_{\a\b}q_\n]
\h^\b(\mathbf{p} ,\lambda ),\nonumber\\
 P^\m=p^\m +\left(p'\right) ^\m, &\quad& q^\m=\left( p ^\prime\right)^\m-p^\m.
\end{eqnarray}
Instead,  the Proca $(P)$ current resulting from the Lagrangian in (\ref{Proca})  reads
\aeq\label{gordonp}
\left( j^\m\right) ^P(\mathbf{p}',\l';\mathbf{p},\l)
=-e \,[\h^{\a}(\mathbf{p}',\l')]^*
[P^\m g_{\a\b}+i[M^{\m\n}_{(1/2,1/2)}]_{\a\b}q_\n]
\h^\b(\mathbf{p},\l).
\ceq
Comparing with Eq.(\ref{gordonv}) we notice  that Proca's theory comes with a previously  built in gyromagnetic factor of $g_V=1$. Instead,   in (\ref{gordonv}) the $g_V$
parameter, that plays the role of the gyromagnetic ratio, is kept free until being fixed  by dynamical constraints such as the unitarity of the Compton scattering amplitudes \cite{Napsuciale:2007ry}. The indicated  discrepancy between the two currents is at the root of the nonequivalence of their respective multipole moments. The latter are usually defined in terms of the electromagnetic  form factors, $F_i$.

The electromagnetic form factors  parametrize the most general third order in the momenta current compatible with Lorentz invariance, which is given by \cite{Lorce:2009bs}
\begin{equation}
j_\mu({\mathbf p}',\lambda';\lambda,{\mathbf p})
=-e\,[\eta^\a({\mathbf p}',\l')]^*
\mathcal{V}_{\alpha\beta\mu}(p',p;F_1,F_2,F_3)\eta^\beta({\mathbf p},\lambda),
\end{equation}
where
\begin{equation}\label{cedc}
\mathcal{V}_{\alpha\beta\mu}(p',p;F_1,F_2,F_3)=g_{\alpha\beta}P_\mu F_1
+(g_{\mu\beta}q_{\alpha}-g_{\alpha\mu}q_\beta)F_2-\frac{1}{2m^2}P_\mu
q_\alpha q_\beta F_3.
\end{equation}
Again, because on mass shell,  $\eta^\alpha({\mathbf p},\l ) p_\alpha=0$, holds valid, the previous expression for the general vertex reduces to
\begin{equation}\label{cedcr}
\mathcal{V}_{\alpha\beta\mu}(p',p;F_1,F_2,F_3)=g_{\alpha\beta}P_\mu F_1
-(g_{\mu\beta}p_{\alpha}+g_{\alpha\mu}p'_\beta)F_2+\frac{1}{2m^2}P_\mu
p_\alpha p'_\beta F_3.
\end{equation}
Performing the multipole decomposition of the latter current is standard. Along the line of Ref.~\cite{Kleefeld:2000nv}, one first calculates in the Breit frame the corresponding charge [$\rho_E({\mathbf q},\l )$]  and current [$\varrho_M(\mathbf{q},\lambda)$] densities, in terms of which the static multipole  moments are defined as
\begin{eqnarray}
\varrho_E(\mathbf{q},\pm 1)&=&
\frac{e\left(\omega ^2-q_3^2\right)}{4 m^2}F_1
+\frac{e \left(4 m^2+q_3^2-\omega ^2\right)}{4 m^2}F_2
-\frac{e\, \omega ^2 \left(4 m^2+q_3^2-\omega ^2\right)}{16 m^4}F_3,\\
\varrho_E(\mathbf{q},0)&=&\frac{e\left(2 m^2+q_3^2\right)}{2 m^2}F_1
-\frac{e\,q_3^2}{2 m^2}F_2
+\frac{e\,q_3^2\,\omega ^2}{8 m^4}F_3,\\
\varrho_M(\mathbf{q},\lambda)&=&-\frac{i\,\lambda\,e\, q_3}{m}F_2, \quad
q=(\omega, {\mathbf q}),
\end{eqnarray}
with $\lambda=-1,0,1$, and $\omega^2=4m^2+\mathbf{q}^2$. Then the related electromagnetic multipole moments are calculated as
\begin{eqnarray}
Q_{E}^l(\lambda)&=&\left.b^{l0}(-i\partial_{\mathbf{q}})\varrho_E(\mathbf{q},\lambda)\right\vert_{\mathbf{q}=0},\\
Q_{M}^l(\lambda)&=&\frac{1}{l+1}\left.b^{l0}(-i\partial_{\mathbf{q}}){\varrho_M}(\mathbf{q},\lambda)\right\vert_{\mathbf{q}=0},
\end{eqnarray}
with the $b^{l0}$ coefficients obtained from the spherical harmonics as
\begin{equation}
b^{l0}(\mathbf{r})=l !\sqrt{{4\pi}/(2l+1)}r^l Y_{l0}(\Omega).
\end{equation}
In effect, in momentum space one encounters
\begin{eqnarray}
b^{00}(-i\partial_{\mathbf{q}})&=&1,\\
b^{10}(-i\partial_{\mathbf{q}})&=&-i\frac{\partial}{\partial q_3},\\
b^{20}(-i\partial_{\mathbf{q}})&=&\frac{\partial^2}{\partial
q_1^2}+\frac{\partial^2}{\pd q_2^2}-2\frac{\partial^2}{\partial q_3^2},\\
b^{30}(-i\partial_{\mathbf{q}})&=&-3\,i\,\frac{\partial}{\partial
q_3}\left(3\frac{\partial^2}{\partial q_1^2}+3\frac{\partial^2}{\partial
q_2^2}-2\frac{\partial^2}{\partial q_3^2}\right)\quad {\mathrm {etc}.}
\end{eqnarray}
Thus, the following relationships between the electromagnetic multipole moments (calculated for the maximal $\lambda=+1$ polarization) and the form factors are found:
\begin{subequations}\label{emmc}
\begin{eqnarray}
Q_E^0(\lambda =+1)&=&e\, F_1,\\
Q_M^1(\lambda =+1)&=&\frac{e\,F_2}{2m},\\
Q_E^2(\lambda =+1) &=&\frac{e}{m^2}(F_1-F_2+F_3),
\end{eqnarray}
\end{subequations}
Here, $Q_E^0(\lambda =+1)$, $Q_M^1(\lambda =+1)$, and $Q_E^2(\lambda =+1)$ are the electric-charge,
magnetic-dipole,  and electric quadrupole moments, respectively. In the following we shall suppress 
the $\lambda $ dependence of the multipoles for the sake of simplifying notations. Then, the following  
relationship  between the  magnetic-dipole,  and the electric quadrupole moments emerges:
\begin{equation}
Q_E^2=\frac{e}{m^2}(F_1+F_3)-\frac{2}{m}Q_M^1.
\label{MD_EQ}
\end{equation}
Comparing the currents in \eqref{gordonv} with  \eqref{cedcr} one finds
\begin{equation}
F_1=1, \quad F_2=g_V,\quad F_3=0.
\label{DFF_FVemmv}
\end{equation}
The associated electromagnetic moments are, correspondingly,
\begin{subequations}\label{emmv}
\begin{eqnarray}
Q_E^0&=&e\,,\\
Q_M^1&=&\frac{e\,g_V}{2m},\\
Q_E^2&=&\frac{e(1-g_V)}{m^2},\label{qv}
\end{eqnarray}
\end{subequations}
and the dipole and quadrupole moments relate as
\begin{equation}
Q_E^2 =\frac{e}{m^2}-\frac{2}{m}Q^1_M.
\end{equation}
As already mentioned before, in \cite{Napsuciale:2007ry} the  free parameter $g_V$ appearing in the Poincar\'e covariant projector method, and thereby $F_2$, has been fixed to the universal value of $F_2=g_V=2$ from the requirement of unitarity of the Compton scattering. With that value, all the multipole moments of the $W$ bosons are correctly reproduced. Compared to this, the Proca current  $j^P_\mu $ in (\ref{gordonp}) has $F_2=1$ which translates to an in-built gyromagnetic ratio that equals one. Because of this difference in the $F_2$ values, the quadrupole moment of the Proca current identically vanishes, and its  magnetic dipole moment is half the one following from (\ref{gordonp}).

This example reveals the nonequivalence of the two methods under consideration and convincingly shows that electromagnetic properties of physical spin-$1$ particles in $(1/2,1/2)$, such as the $W$ boson, are adequately described by means of the Lagrangian in (\ref{PCPF}). The gauged Lagrangian ${\mathcal L}_P$ in (\ref{Proca}) is incomplete in the sense that it describes spin-$1$ particles with vanishing quadrupole moments.

The nonuniqueness discussed above does not restrict to a given representation alone. It turns out that different representations containing spin $1$ can also provide wave
equations equivalent to (\ref{prowafu}) and (\ref{procaext}) at the noninteracting level. To illuminate this point, in the next subsection we present the description of spin $1$ by means of the antisymmetric Lorentz tensor of second rank.

\subsection{\label{s2b}Nonuniqueness of the spin-1 description in \texorpdfstring{$(1,0)\oplus(0,1)$}\space{}}

Alternatively, one could consider spin $1$ transforming in the pure spin $1$ Lorentz-algebra representation, $(1,0)\oplus (0,1)$, whose 6 degrees of freedom  can be mapped onto an antisymmetric Lorentz tensor of second rank, and denoted by $w^{\alpha\beta}({\mathbf p},\l)$ in momentum space.
These 6 degrees of freedom are distributed over a pair of spin-$1$ bosons of negative and positive parities, in turn denoted by $w^{\alpha\beta}_+({\mathbf p},\lambda)$, and $w^{\alpha\beta}_-({\mathbf p},\lambda)$ with $\l =-1,0,1$ (see \cite{DelgadoAcosta:2012yc} for technical details); they are related as 
\aeq
w^{\alpha\beta}_+({\mathbf p},\lambda)=\c^{\a\b}{}_{\s\r}w^{\s\r}_-({\mathbf p},\lambda),
\ceq
where 
\aeq\label{chi}
\c^{\a\b\g\d}=\frac{i}{2}\e^{\a\b\g\d}
\ceq
is the chirality operator in this representation.
\subsubsection{\label{nproca}Deriving a Proca's wave equation from a combined mass-\texorpdfstring{$m$}\space{} and negative-parity projector: Causality test}

Now the projector operator in (\ref{waqu_Prjt}) can be chosen in parallel to (\ref{fourvprj}) as 
\begin{equation}
\Pi_{(1,0)\oplus (0,1)}^{\a\b\g\d}({\mathbf p})=\frac{{\widehat P}^2}{m^2}{\mathbf
P}^{\a\b\g\d}_\pm ({\mathbf p}).
\label{proj_AT}
\end{equation}
The momentum-space operators ${\mathbf P}_+^{\a\b\g\d}({\mathbf p})$ and ${\mathbf P}_-^{\a\b\g\d}({\mathbf p})$, in turn  project on the respective
amplitudes,  $w_+^{\alpha \beta }({\mathbf p}, \l)$ and $w_-^{\alpha \beta}({\mathbf p}, \l)$, of the positive and negative-parity  states of the quantum $B^{\a\b}(x)$field in $\tensorrep$ (the explicit expressions for them can be consulted in \cite{DelgadoAcosta:2012yc}). They are defined as
\aec
\mathbf{P}_-^{\a\b\g\d}({\mathbf p})
&=&\sum_{\l}w_-^{\a\b}({\mathbf p},\l)[w_-^{\g\d}({\mathbf p},\l)]^*\nonumber\\
&=&\frac{1}{2\,p^2}\(-p^\b p^\g g^{\a\d}-p^\d p^\a g^{\b\g}+g^{\b\d}p^\a p^\g
+g^{\a\g} p^\b p^\d\),
\label{par_T_plus}\\
\mathbf{P}_+^{\a\b\g\d}({\mathbf
p})&=&\mathbf{1}^{\a\b\g\d}-\mathbf{P}_-^{\a\b\g\d}({\mathbf
p}),\label{par_T_minus}\\
{\mathbf
1}^{\a\b\g\d}&=&\frac{1}{2}(g^{\a\g}g^{\b\d}-g^{\a\d}g^{\b\g})\label{aunit}.
\cec

Then the corresponding wave equations read
\begin{subequations}\label{Gl1}
\aec
\frac{{\widehat P}^2}{m^2}\mathbf{P}_{\pm} ^{a\b\g\d}(p)[w_{\pm }({\mathbf
p},\l)]_{\g\d}
&=&w_\pm ^{\a\b}({\mathbf p},\l),\\
~[-m^2\mathbf{1}^{\a\b\g\d}+p^2\mathbf{P}_{\pm}
^{\a\b\g\d}(p)][w_\pm ({\mathbf p},\l)]_{\g\d }&=&0.
\cec
\end{subequations}
The momentum space propagators associated with Eq.~ (\ref{Gl1}) are found as
\aeq
\[-m^2\mathbf{1}^{\a\b\s\r}+p^2\mathbf{P}_{\pm}^{\a\b\s\r}(p)\]\[S_{\pm}(p)\]_{\s\r}{}^{\g\d}
=\mathbf{1}^{\a\b\g\d},
\ceq
and amount to
\begin{subequations}\label{pmprop}
\aec
~[S_{\pm}(p)]^{\a\b\g\d}&=&\[-m^2\mathbf{1}^{\a\b\g\d}+p^2\mathbf{P}_{\pm}
^{\a\b\g\d}(p)\]^{-1}\\
&=&\frac{1}{p^2-m^2+i\,\e}
\[\frac{p^2}{m^2}\mathbf{P}_{\pm}
^{\a\b\g\d}(p)-\frac{p^2-m^2}{m^2}\mathbf{1}^{\a\b\g\d}\].
\cec
\end{subequations}
The explicit forms of (\ref{Gl1}) in configuration space are then obtained as
\aec
~\[1_{\a\b\g\d}m^2+\frac{1}{2}\(g_{\a\g}\pd_{\b}\pd_\d -g_{\a\d}\pd_\b \pd_\g -g_{\b\g}\pd_{\a}\pd_{\d}+g_{\b\d}\pd_\a \pd_\g \)\]B_-^{\g\d}&=&0,
\label{AT_minus}\\
~\[1_{\a\b\g\d}(\pd^2+m^2)-\frac{1}{2}\(g_{\a\g}\pd_{\d}\pd_\b -g_{\a\d}\pd_\g \pd_\b -g_{\b\g}\pd_{\d}\pd_{\a}+g_{\b\d}\pd_\g \pd_\a \)\]B_+^{\g\d}&=&0,
\label{AT_plus}
\cec
where  $B_+^{\a\b}(x,\l) $ and $B_-^{\a\b} (x,\l) $ (arguments suppressed)
are the Fourier transforms of $w_+^{\a\b}({\mathbf p},\l )$, and
$w_+^{\a\b}({\mathbf p},\l )$, respectively.

Equation (\ref{AT_minus}) does not contain a $ \partial^2$ term,  in parallel to the nonpropagating scalar sector of the four-vector in (\ref{nonprop_1}).
Instead, the wave equation for the positive parities of the antisymmetric tensor  parallels the wave equation for the negative-parity sector of the four-vector in (\ref{parpro_proca}). It is now instructive to compare the above approach to Proca's theory where a freely moving antisymmetric-tensor field is described by the following  textbook Lagrangian:
\begin{equation}
{\mathcal L}_T^{(P)}=\frac{1}{4}m^2B_{\mu\nu}B^{\mu\nu}-\frac{1}{2}G_\mu G^\mu,
\quad G_\mu =\partial ^\alpha B_{\alpha \mu}.
\label{Tensor1}
\end{equation}
Then, the  corresponding Euler-Lagrange equation emerges as
\begin{eqnarray}
\partial _\alpha \partial^\delta B_{\delta \beta} -\partial _\beta
\partial^\mu B_{\mu\alpha} &=&-m^2 B_{\a\b}.
\label{tensorwafu1}
\end{eqnarray}
A comparison of (\ref{tensorwafu1}) with (\ref{AT_minus}) reveals their identity for the noninteracting case, in which the anti-commutators of the ordinary derivatives are vanishing. Upon gauging, these commutators will give rise to the electromagnetic field tensor and to nongradiental Yukawa tensor-tensor couplings.
 
Also here we shall test the hyperbolicity and causality of the gauged equations (\ref{AT_plus}), on the one side, and of (\ref{tensorwafu1}) and (\ref{AT_minus}), on the other, that incorporate the following  gauged auxiliary conditions (obtained by contracting the respective equations be $D_\alpha$): 
\begin{equation}
-m^2 D_\sigma B_+^{\sigma\rho}= \frac{ie}{2}
\left[ -F_{\kappa \tau}D^\rho +g^\rho \, _\kappa F^\sigma{}_\tau D_\sigma
-g^\rho{} _\tau F^\sigma{}_\kappa D_\sigma\right]B^{\kappa\tau}_+,
\label{pospar_auxcond}
\end{equation}
for the positive  parities, and
\begin{eqnarray}\label{negpar_auxcond}
-m^2 D_\s B_-^{\s\r}
=\frac{ie}{2}\[F_{\kappa}{}^\r D_\t -F_\t{}^\r D_\kappa+D^\r
F_{\k\t}\]B_-^{\kappa\t}+D^2 D_\kappa B_-^{\kappa\r}
\end{eqnarray}
for negative ones. With that the gauged equations (\ref{AT_plus}) [equivalent to(\ref{tensorwafu1})] and (\ref{AT_minus}) become
\begin{equation}
\label{AS_Pr_PosPar}
~\[1_{\a\b\g\d}(D^2+m^2)-2ie\mathbf{1}_{\a\b}{}^{\s\r}F^\m{}_\r
\mathbf{1}_{\s\m\g\d}
-\frac{ie}{m^2}\mathbf{1}_{\a\b}{}^{\m\n}D_\m F_{\g\d}D_\n
-\frac{2ie}{m^2}\mathbf{1}_{\a\b}{}^{\m\n}D_\m F_{\s\r}g_{\n\t}D^\s
\mathbf{1}^{\r\t}{}_{\g\d}
\]B^{\g\d}_+=0,
\end{equation}
\begin{equation}\label{AS_Pr_NegPar}
\[1_{\a\b\g\d}m^2+\frac{2}{m^2}\(\frac{1}{4}e^2 F_{\a\b} F_{\g\d}
+\mathbf{1}_{\a\b}{}^{\m\n}D_\m
(ieF_{\n\s}+g_{\n\s}D^2)D_\r\mathbf{1}^{\s\r}{}_{\g\d}\)\]B_-^{\g\d}=0.
\end{equation}
The latter equations show that while the negative-parity equation is of fourth order, the positive-parity one conserves its second order, meaning that the hyperbolicity of the negative-parity equation remains inconclusive at this stage because the Currant-Hilbert criterion solely applies to equations of maximally second order. This behavior is quite unexpected indeed as it resembles more  the nonpropagating scalar sector in the four-vector, whose genuine gauged equation (\ref{bad_gg}) is of fourth order, than the well-behaved polar vector sector in (\ref{gut_gg}). Therefore,  only the hyperbolicity of the positive parity equation can be tested. In so doing, the characteristic determinant is straightforwardly calculated as $n^{12}$, similar to (\ref{Proca_chdt}), thus implying both  hyperbolicity and causality of this very equation. Therefore, it is the positive parity sector in $(1,0)\oplus (0,1)$ that better matches the negative-parity sector in $(1/2,1/2)$. This situation is alerting about any precipitate conclusions on the equivalent descriptions of spin $1$ by means of  $(1/2,1/2)$ and $(1,0)\oplus(0,1)$. Our next point throws more light on this issue.

\subsubsection{The vector-tensor equivalence theorem within Proca's approach}

In contracting Eq. (\ref{tensorwafu1}) from above  by $\pd^\alpha$ amounts to
\aec
\pd^2 G_\b-\pd^\a\pd_\b G_\a +m^2 G_\b=(\pd^2+m^2)G_\b-\pd^\a\pd_\b G_\a&=&0,\quad{ G}_\eta =\partial^\gamma B_{\gamma \eta}.\label{tensorwafu2}
\cec

In now rescaling  ${ G}_\eta$ by an inverse mass with the purpose of matching dimensionality, i.e. in introducing $V_\eta =\frac{1}{m^2}{ G}_\eta$,  Eq. (\ref{tensorwafu2}) becomes
\begin{eqnarray}
(\partial ^2 +m^2 )V_\beta -\pd^\a \pd_\b V_\a&=& 0\label{tensorwafu},
\end{eqnarray}
which is equivalent to (\ref{prowafu}). Both equations lead to equivalent theories for noninteracting spin-$1$ particles. The above equivalence reflects the fact that any
antisymmetric tensor of second rank is representable in terms of a four vector, $V_\m$, according to
\begin{equation}
B_{\mu\nu}=\partial _\m V_\nu -\partial _\nu V_\mu,
\label{vct_tnsr_equiv}
\end{equation}
meaning that the knowledge on $B_{\mu\nu}$ is reduced to the knowledge on $V_\mu$. Therefore, the (four-vector)-(antisymmetric tensor) equivalence theorem (shortly, the vector-tensor tensor equivalence) \cite{Jenkins:1972pd} concerns, in the first place the solutions of the wave equations (\ref{prowafu}) and (\ref{tensorwafu}). However, it should be  obvious that the equivalence extends to the genuine gauged equations too in complete parallel to Eqs.~(\ref{parpro_proca}) and (\ref{gut_gg}) from above. Below we shall see that this equivalence also fully extends to the electromagnetic currents, having its validity for the particular case of a gyromagnetic ratio taking the unphysical unit value. Beyond this, and for the case of the general Poincar\'e covariant projector method, the equivalence will be removed.

\subsubsection{Spin-\texorpdfstring{$1$}\space{} wave equation from the
Poincar\'e covariant projector method: Causality proof}
The counterpart of Eq. (\ref{Prime_one}) for the antisymmetric tensor,
\begin{equation}
\Pi^{\a\b\g\d }({\mathbf p})_{(1,0)\oplus (0,1)}w_{\g\d}({\mathbf p},\l
)=w_{\g\d}({\mathbf p},\l ),
\quad \Pi^{\a\b\g\d }({\mathbf p})_{(1,0)\oplus (0,1)}=-\frac{1}{2m^2}\left[
W^2_{(1,0)\oplus (0,1)}({\mathbf p})\right] ^{\a\b\g\d},
\label{Lub_Ten}
\end{equation}
can be elaborated too, and same argument as the one that conduced to  Eq.~(\ref{procaext}) can also be applied to extend (\ref{AT_plus}) toward a Lagrangian which contains the most general antisymmetric derivative contributions. Such an extension has been performed within the Poincar\'e covariant projector method and the following result has been found  in \cite{DelgadoAcosta:2012yc}:
\aeq\label{eomva}
~\[\G_{\a\b\g\d\m\n} p^{\mu}p^{\nu}-m^{2}\mathbf{1}_{\a\b\g\d}\]  w^{\g\d}
({\mathbf p},\l)=0,
\ceq
with the antisymmetric unit ${\mathbf 1}_{\a\b\g\d}$ as defined in
\eqref{aunit} and
\begin{eqnarray}
\label{Tmntensor}
\Gamma_{\alpha\beta\gamma\delta \mu\nu}&\,=\,&g_{\mu\nu}{\mathbf
1}_{\alpha\beta\gamma\delta}-i {\widetilde g}
[\mathcal{M}^{T}_{\mu\nu}]_{\alpha\beta\gamma\delta},\\
\left[\left(
\mathcal{M}^{T}\right)^{\mu\nu}\right]_{\alpha\beta\gamma\delta}&\,=\,&\frac{1}{2}\left[
(M^{\mu\nu}_{(1/2,1/2)})_{\alpha\gamma}g_{\beta\delta}
+g_{\alpha\gamma}(M_{(1/2,1/2)}^{\mu\nu})_{\beta\delta}
-(M_{(1/2,1/2)}^{\mu\nu})_{\alpha\delta}g_{\beta\gamma}
-g_{\alpha\delta}(M_{(1/2,1/2)}^{\mu\nu})_{\beta\gamma}
\right].\label{gentensor}
\end{eqnarray}
Here,  ${\mathcal M}^T_{\mu\nu}$ stand for the Lorentz-algebra generators in the antisymmetric tensor representation, and  ${\widetilde g}$ is again a free parameter, to be associated in the following with the gyromagnetic ratio in $(1,0)\oplus(0,1)$. Equation (\ref{eomva}) is satisfied by both parities, a reason for which no parity index, $\pm$, has been attached to the solutions. In position space, the equation of motion (\ref{eomva}) can be expressed more transparently as 
\begin{equation}
\left( \partial^2 +m^2\right)B_{\alpha \beta}  +
\frac{\widetilde g}{2}\left( \left[ \partial_\alpha,\partial_ \gamma\right]
B^\gamma{}_\beta
+\left[ \partial _\beta,\partial _\delta \right] B^{\delta}{}_{\alpha}
-\left[ \partial _\alpha,\partial _\delta\right]  B_\beta{}^{\, \delta}
-\left[ \partial _\beta,\partial _\gamma\right]  B _\alpha{}^{ \gamma} \right) =0.
\label{tensorwafuext}
\end{equation}
The former equation follows from the Lagrangian \cite{DelgadoAcosta:2012yc}:
\begin{eqnarray}
{\mathcal L}_T(x)=\left(  \partial^{\mu}B_{\alpha\beta}(x)\right)^{*}
\Gamma^{\alpha\beta}_{\gamma\delta\mu\nu} \partial ^{\nu}B^{\gamma\delta}(x)
&-&m^{2}\left(B^{\alpha\beta}(x)\right)^{*}B_{\alpha\beta}(x).
\label{lag1001b}
\end{eqnarray}
Comparing  (\ref{tensorwafuext}) to (\ref{AT_plus}) one sees that they are equivalent for free particles though, in accounting for (\ref{gauging}), they are quite different upon gauging. The gauged Lagrangian corresponding to (\ref{lag1001b}) reads
\begin{eqnarray}
\mathcal{L}_T(x)=\left(  D^{\mu}B_{\a\b}(x)\right)^{*}\G^{\a\b}_{\g\d\m\n}
D^{\nu}B^{\g\d}(x)
&-&m^{2}\left(B^{\a\b}(x)\right)^{*}B_{\a\b}(x).
\label{lag1001gauged}
\end{eqnarray}
That $\mathcal{L}_T(x)$ is Hermitian can be verified using
\aeq\label{gh}
\G^{*}_{\a\b\g\d\m\n}=\G_{\a\b\g\d\m\n}=\G_{\g\d\a\b\n\m}.
\ceq
The propagator associated to (\ref{eomva}) reads
\aec\label{tprop}
[S(p)]^{\a\b\g\d}=[\G^{\a\b\g\d}p_\m p_\n-m^2 \mathbf{1}^{\a\b\g\d}]^{-1}
&=&\frac{\mathbf{1}^{\a\b\g\d}}{p^2-m^2+i\,\e}.
\cec
The proof of the hyperbolicity of Eq. (\ref{tensorwafuext}) is quite easy. Indeed, in noticing that upon gauging the derivatives entering the commutators will give rise to the electromagnetic field tensor, one immediately sees that the only second order derivatives entering the characteristic determinant are those contained in the diagonal kinetic term. In this way  a genuine differential equation is obtained without  any need for invoking supplementary conditions. Then the characteristic determinant  simply equals $(n^2)^6$, thus making the hyperbolicity and causality in question  manifest.
 
Finally, it has to be emphasized that differently from Proca's single-parity equation in (\ref{AS_Pr_NegPar}) and its inconclusive hyperbolicity, the Poincar\'e
covariant projector method describes all 6 degrees of freedom constituting the spin-$1$ parity doublet as causally propagating, a circumstance that will acquire
significant  importance in Sec. \ref{sec3b} below in connection with obtaining at high energies finite differential and total  cross sections for Compton scattering off $(1,0)\oplus (0,1)$.

\subsection{Representation dependence of the  currents: Violation of the vector-tensor equivalence theorem for \texorpdfstring{${\widetilde g}\not=1$}\space{}}
\subsubsection{Electromagnetic current for \texorpdfstring{$\tensorrep$}\space{} from mass-\texorpdfstring{$m$}\space{} and negative-parity projector methods}
In order to find the Noether current of $(1,0)\oplus (0,1)$ bosons within  the combined mass and parity-projector methods it is useful to rewrite \eqref{Gl1} as
\aeq
\[T^-_{\a\b\g\d\m\n}p^\m p^\n-m^2\mathbf{1}_{\a\b\g\d}\]w_-^{\g\d}=0,
\ceq
where
\aec\label{tmprocat}
T^-_{\a\b\g\d\m\n}&=&-\frac{1}{2}[(M_{(1/2,1/2)})_{\m\s}]_{\a\b}[(M_{(1/2,1/2)})_{\n\r}]_{\g\d} g^{\s\r}\\
&=&-\frac{1}{2}[g_{\a\m}(g_{\b\g}g_{\d\n}-g_{\b\d}g_{\g\n})+g_{\b\m}(g_{\a\d}g_{\g\n}-g_{\a\g}
g_{\d\n})].
\cec
The corresponding electromagnetic current involving the negative-parity states is then given by
\aeq
j_\m^{-}(p',\l';p,\l)=
- e\,[w^{\alpha\beta}_-(p',\lambda')]^*
\mathcal{O}_{\alpha\beta\gamma\delta\mu}^{(-)}(p',p)
w^{\gamma\delta}_-(p,\lambda)
\label{NPRTY_AT}
\ceq
with
\aeq
\mathcal{O}_{\alpha\beta\gamma\delta\mu}^{-}(p',p)=T^{-}_{\a\b\g\d\n\m}p'^\n+T^{-}_{\a\b\g\d\m\n}p^\n
.
\ceq
This vertex satisfies the Ward-Takashi identity,
\aeq
(p'-p)^\m\mathcal{O}_{\alpha\beta\gamma\delta\mu}^{-}(p',p)=[S^{-}_{\a\b\g\d}(p')]^{-1}-[S^{-}_{\a\b\g\d}(p)]^{-1},
\ceq
with the propagator defined in \eqref{pmprop}. In making use of the standard representation of the antisymmetric tensor in terms of the four-vector,
\aec
w^{\a\b}_-(p,\l)&=&\frac{1}{\sqrt{2}m}\[p^a \h^\b(p,\l)-p^\b \h^\a(p,\l)\]
=\frac{-i}{\sqrt{2}\,m} [M_{(1/2,1/2)}^{\m\n}]^{\a\b}p_\m \h_\n(p,\l),
\cec
\{with a normalization factor of $1/(\sqrt{2} m)$, so that $[w_-^{\a\b}(p,\l)]^* [w_-(p,\l)]_{\a\b}=-1$\} allows us to reexpress (\ref{NPRTY_AT}) as a current between four-vector states according to
\begin{eqnarray}
j_\mu^{-}(p',\lambda';p,\lambda)
&=&- e\,[\eta^\alpha(p',\lambda')]^*
\mathcal{V}^{-}_{\alpha\beta\mu}(p',p)\eta^\beta(p,\lambda).
\label{NPRTY_ATFV}
\end{eqnarray}
As long as the $\h^\a(p)$ states are on mass shell, the terms in $\mathcal{V}^{-}_{\alpha\beta\mu}(p',p)$ containing  $p'^\a$ and $p^\b$ can be omitted, which leads to
\aeq
\mathcal{V}^{-}_{\alpha\beta\mu}(p',p)=(p'+p)_\m g_{\a\b}-(p_\a g_{\b\m}-p'_\b g_{\a\m}).
\ceq
Comparing the latter expression with the general vertex (\ref{cedcr}), we
find the
following parametrization for this current:
\begin{align}
F_1= 1,\qquad F_2=1, \qquad F_3=0.
\label{Procas_ELMFF}
\end{align}
One sees  that the electromagnetic multipole moments,
\begin{equation}
Q_E^0=e, \quad Q_M^1=\frac{e}{2m}, \quad Q_E^2=0,
\label{P_junk}
\end{equation}
are same as those of a Proca particle. In consequence one observes that for the particular case of Proca's approach, due to the in-built $g_V=1$ value, the vector-tensor equivalence theorem  extends to the level of the respective electromagnetic currents. However, as long as the electromagnetic multipole moments in (\ref{P_junk}) are unphysical, the equivalence theorem cannot be given a physical status either. Rather, one has to seek to confirm it independently within methods known for their physical predictive power, among them the Poincar\'e covariant projector formalism. Below we shall see that the vector-tensor equivalence is not universal but can appear violated at the level of the currents which, in general, turn out to be characterized by different form factors.

\subsubsection{Electromagnetic current for \texorpdfstring{$\tensorrep$}\space{}
within the Poincar\'e covariant projector method{}}
The Lagrangian in (\ref{lag1001gauged}) gives rise to the following electromagnetic current in momentum space:
\begin{equation}
j^{T}_{\mu}(\mathbf{p},\l;\mathbf{p}',\l')
=-e [{w}^{\a\b}(\mathbf{p}',\l')]^*
\[\G_{\a\b\g\d\n\m}p'^\n+\G_{\a\b\g\d\m\n}p^\n\]
w^{\g\d}(\mathbf{p},\l).
\label{currentbv}
\end{equation}
{}From now onwards  we restrict to the vertex that is of first order in the charge of the particle,
\aeq\label{vdef}
V(p',p)_{\a\b\g\d\m}=\G_{\a\b\g\d\n\m}p'^\n+\G_{\a\b\g\d\m\n}p^\n;
\ceq
this vertex satisfies the Ward-Takahashi identity,
\aeq\label{wtnkr}
(p'-p)^\m V(p',p)_{\a\b\g\d\m}=[S(p')_{\a\b\g\d}]^{-1}-[S(p)_{\a\b\g\d}]^{-1},
\ceq
with the propagator defined in \eqref{tprop}. If we express the $\G_{\a\b\g\d\m\n}$ tensor in terms of the $[\mathcal{M}^T_{\m\n}]_{\a\b\g\d}$ generators in (\ref{gentensor}), the following two-term convection plus spin-magnetization current emerges:
\begin{equation}
j^{T}_{\mu}(\mathbf{p},\l;\mathbf{p}', \l')
=-e[{w}^{\a\b}(\mathbf{p}',\l')]^*\[ P_{\m}\mathbf{1}_{\a\b\g\d}
+i{\widetilde g}
\left(\mathcal{M}^T_{\m\n}\)_{\a\b\g\d}q^{\nu}\]w^{\g\d}(\mathbf{p},\l).
\label{gordonbv}
\end{equation}

\noindent
It is now quite instructive to reexpress this current in terms of the four-vector degrees of freedom in reference to (\ref{vct_tnsr_equiv}) and using $\eta ({\mathbf p},\lambda)$ from (\ref{curr1}). In so doing, the following equality is found:
\begin{eqnarray}\label{ctrans}
\left( j^{T}\right)^{\mu}(\mathbf{p},\l;\mathbf{p}',\l')
&=&-e[\h^{\a}(\mathbf{p}',\l')]^*
\left\lbrace P^{\m}g_{\a\b}\,{\widetilde g}+i{\widetilde g}
\left[M_{(1/2,1/2)}^{\m\n}\]_{\a\b}q_{\nu}\right.\nonumber\\
&&\left.+\frac{e }{2m^2}2({\widetilde g}-1)
\({p_\alpha p'_\beta}
-{(p'\cdot p)}g_{\alpha\beta} \)P^\mu\right\rbrace
\h^{\b}(\mathbf{p},\l).
\end{eqnarray}
The vertex, $\mathcal{V}^{T}_{\alpha\beta\mu}(p',p)$, corresponding to this current is given by the expression in the curly brackets and reduces on shell to
\aec
\mathcal{V}^{T}_{\alpha\beta\mu}(p',p)&=&g_{\alpha\beta}P_\mu {\widetilde g}
-(g_{\mu\beta}p_\alpha+g_{\alpha\mu}p'_\beta){\widetilde g}
+\frac{1}{m^2}P_\mu p_\alpha p'_\beta ({\widetilde g}-1)\nonumber\\
&&+\frac{1}{m^2}g_{\alpha\beta}P_\mu (p'\cdot p)(1-{\widetilde g}),
\cec
Notice that except  for the particular ${\widetilde g}=1$ value, the latter set is distinct from Proca's form factors in (\ref{Procas_ELMFF}).

A direct comparison of the latter equation with (\ref{cedcr}) allows us to identify the electromagnetic form factors as 
\begin{equation}\label{parametersva}
F_1={\widetilde g}+\frac{1}{m^2}(1-{\widetilde g})(p'\cdot p), \quad F_2={\widetilde g}, \quad F_3=-2(1-{\widetilde g}).
\end{equation}
Here, a momentum dependence in $F_1$  had to be included in order to account for the extension of the current $\left(j^V\right)^\mu ({\mathbf p },\l;{\mathbf p}^\prime,\l^\prime)$ in (\ref{gordonv}) to the current $\left(j^{T}\right)^{\mu}(\mathbf{p},\l;\mathbf{p}',\l')$ in (\ref{ctrans}). In the Breit frame we have 
$p'\cdot p=m^2+\mathbf{q}^2/2$; taking into account this $\mathbf{q}^2$ dependence in $F_1(\mathbf{q}^2)$, the calculation of the corresponding multipole moments gives a similar result to \eqref{emmc} but with the replacement $F_1\rightarrow F_1(0)=1$. For a particle of $\lambda=1$ we get
\begin{subequations}\label{emmt}
\begin{eqnarray}
Q_E^0&=&e\,\\
Q_M^1&=&\frac{e\,{\widetilde g}}{2m},\\
Q_E^2&=&-\frac{e(1-{\widetilde g})}{m^2}.
\end{eqnarray}
\end{subequations}
Comparing the last equations with (\ref{emmv}), one immediately realizes that for ${\widetilde g}=g_V\not=1$ the sign of the quadrupole moment in the antisymmetric-tensor representation is inverted relative to the one observed in the four-vector. Our observation is in accord with a similar finding reported  earlier in \cite{Shay:1968iq}. We here attribute this  peculiarity  to the different  $F_i$ values taken in the four-vector (\ref{DFF_FVemmv}), in comparison with those characterizing the 
antisymmetric-tensor (\ref{parametersva}) representations (again, for ${\widetilde g}=g_V\not=1$). To be specific, the first impression is that the two sets of form factors under discussion are characterized by two different $F_1$ and $F_3$ values. However, as long as $F_1(\mathbf{q}^2)$ enters the multipole moments as $F_1(0)=1$, its contribution to these observables is effectively equivalent to its counterpart $F_1=1$ within the four-vector. In effect, one is left with  the difference in $F_3$ and it is that very form factor that provokes the aforementioned change of sign in $Q^2_E$.

The above considerations show beyond any doubt that within the Poincar\'e covariant projector method the current in the antisymmetric tensor in (\ref{gordonbv}) and its equivalent in (\ref{ctrans}) is quite different from the one corresponding to spin 1  in $(1/2,1/2)$ and given in (\ref{gordonv}).
 
To recapitulate, we presented a total of four different spin-$1$ wave equations, equivalent at the free-particle level, and, in general, distinct upon gauging. Not all of them predict electromagnetic multipole moments which can be attributed to experimentally observed particles. The predictions of the Poincar\'e covariant projector method for the multipole moments of a fundamental boson transforming as a four-vector \cite{Napsuciale:2007ry} have successfully  matched data on the $W$ boson. As to the multipole moments calculated for the antisymmetric-tensor representation, no certain match with observed mesons can be claimed so far, mainly because of the composite nature of hadrons. Recent lattice QCD simulations \cite{Hedditch:2007ex} do indeed predict an oblate $\rho$ meson quadrupole moment, just as is known for the $W$ boson, which seems to bring the traditional candidates for $(1,0)\oplus (0,1)$ residents, the $\rho$- and $a_1$ mesons, closer to the four-vector representation. However, in order to correctly pin down the relevant representation, a  careful and reliable study on the effect of the internal structure on the interpolation of the electromagnetic form factors to zero momentum transfer is indispensable. We here take the position that the unitarity of the Compton scattering cross section in the ultraviolet provides a reliable criterion on the observability of the states in any given Lorentz representation. It is the goal of the present investigation to employ the latter criterion in testing the  credibility of the pure spin-$1$ wave equation following from the Poincar\'e covariant projector method, and analyze the observability of massive fundamental spin-$1$ bosons residing in the antisymmetric tensor. The rest of the paper is devoted to this issue.

\section{Compton scattering in  \texorpdfstring{$(1,0)\oplus (0,1)$}\space{}
within the Poincar\'e covariant projector method}
\label{c1001}
\subsection{General formulas for the scattering amplitudes}
\begin{figure}
\begin{minipage}{0.3\textwidth}
\bigskip
\includegraphics{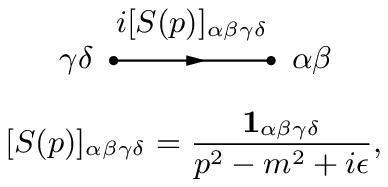}
\end{minipage}
\begin{minipage}{0.3\textwidth}
\bigskip
\begin{center}
\includegraphics{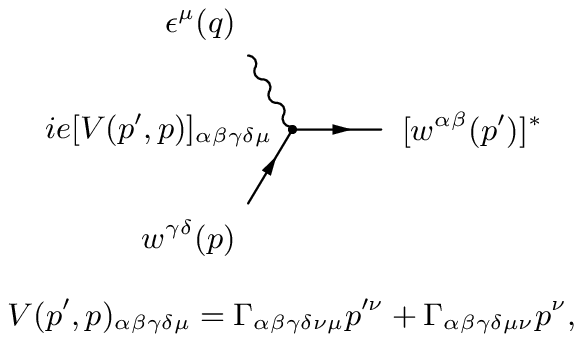}
\end{center}
\end{minipage}
\begin{minipage}{0.3\textwidth}
\bigskip
\includegraphics{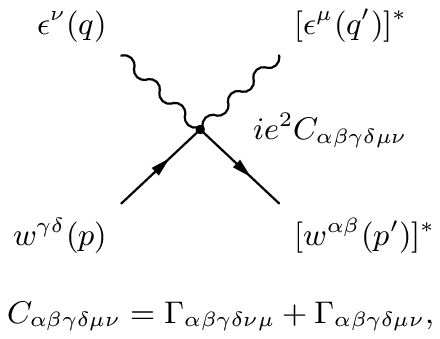}
\end{minipage}
\caption{\label{rules} Feynman rules corresponding to the Lagrangian in
\eqref{lag1001b} in combination with
eqs. \eqref{Tmntensor}-\eqref{gentensor}.}
\end{figure}

\begin{figure}[ht]
\bigskip
\begin{center}
\includegraphics{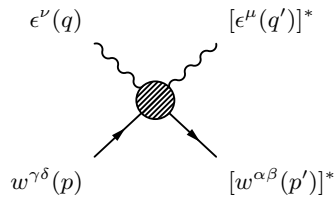}
\end{center}
\caption{\label{cs} Schematic presentation of the process of Compton scattering.}
\end{figure}

The Feynman rules extracted from the interaction Lagrangian (\ref{lag1001b}) together with the propagator calculated as the inverse of the equation of motion (\ref{eomva}) are depicted in Fig. \ref{rules}, while the notations used for the process under investigation are presented in Fig. \ref{cs}.
\begin{figure}
\begin{minipage}{0.3\textwidth}
\bigskip
\begin{center}
\includegraphics{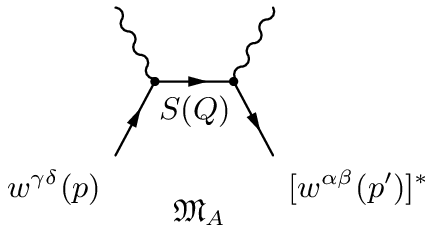}
\end{center}
\end{minipage}
\begin{minipage}{0.3\textwidth}
\bigskip
\begin{center}
\includegraphics{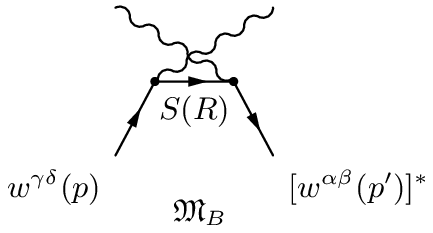}
\end{center}
\end{minipage}
\begin{minipage}{0.3\textwidth}
\bigskip
\begin{center}
\includegraphics{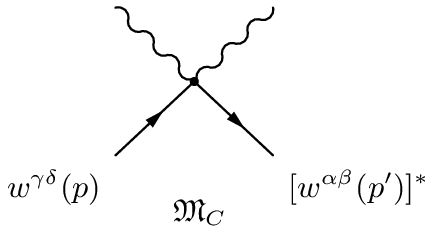}
\end{center}
\end{minipage}
\caption{\label{Ms} Feynman diagrams for the contributions to the tree level
Compton scattering amplitude. For notations see the text.}
\end{figure}
Finally, the scattering amplitude is determined by the contributions of the three diagrams in Fig. \ref{Ms} as
\aeq\label{amplitudet}
\M=\M_A+\M_B+\M_C,
\ceq
the explicit expressions being given by
\begin{align}
i\,\M_A&= e^2[{\ef}^{\a\b}({\mathbf
p}^\prime,\lambda')]^*V{(p',Q)}_{\a\b\h\z\m}\P(Q)^{\h\z\q\f}
V(Q,p)_{\q\f\g\d\n}\ef^{\g\d}({\mathbf p},\lambda )
[\e^{\m}({\mathbf q}^\prime, \ell^\prime )]^*\e^{\n}({\mathbf q},\ell),\\
i\,\M_B&=  e^2[{\ef}^{\a\b}({\mathbf p}^\prime, \lambda^\prime
)]^*V(p',R)_{\a\b\h\z\n}\P(R)^{\h\z\q\f}
V(R,p)_{\q\f\g\d\m}\ef^{\g\d}({\mathbf p},\lambda )
[\e^{\m}({\mathbf q}^\prime, \ell^\prime )]^*\e^{\n}({\mathbf q},\ell),\\
i\,\M_C&= -e^2[{\ef}^{\a\b}({\mathbf p}^\prime, \lambda^\prime)]^*
V_{\a\b\g\d\m\n}\ef^{\g\d}({\mathbf p},\lambda )
[\e^{\m}({\mathbf q}^\prime, \ell^\prime )]^* \e^{\n}({\mathbf q},\ell).
\label{amplitudes}
\end{align}
Here $Q=p+q=p'+q'$ and $R=p'-q=p-q'$ are in turn the momenta of the intermediate states in the $s$ and $u$ channels, $p\, (p')$ stands for the momentum of the incident
(scattered) target boson, while $q\,(q')$ is the momentum of the incident (scattered) photon. The Ward-Takahashi identity \eqref{wtnkr} leads to $\M(\e(q)\rightarrow q)=0$, meaning the gauge invariance of the amplitude. Insertion of  the propagator in (\ref{amplitudes}) amounts to
\aeq\label{amplitude}
i\,\M=e^2[{\ef}^{\a\b}({\mathbf p}^\prime,\lambda^\prime  )]^*
U_{\a\b\g\d\m\n}\ef^{\g\d}({\mathbf p},\l)
[\e^{\m}({\mathbf q}^\prime, \ell^\prime)]^*\e^{\n}({\mathbf q},\ell),
\ceq
with $\epsilon_\mu$ denoting the photon polarization vectors, and
\aeq\label{udef}
U(p',P,Q,p)_{\a\b\g\d\m\n}\equiv\frac{V(p',Q)_{\a\b\q\f\m}V(Q,p)^{\q\f}{}_{\g\d\n}}{Q^2-m^2}
+\frac{V(p',R)_{\a\b\q\f\n}V(R,p)^{\q\f}{}_{\g\d\m}}{R^2-m^2}
-V_{\a\b\g\d\m\n}.
\ceq
The complex conjugate to this amplitude is obtained with the aid of (\ref{gh}), (\ref{vdef}), and (\ref{udef}) as
\begin{align}
V^*(k',k)_{\a\b\g\d\m}=&V(k',k)_{\a\b\g\d\m}=V(k,k')_{\g\d\a\b\m},\\
U^*(p',Q,R,p)_{\a\b\g\d\m\n}=&U(p',Q,R,p)_{\a\b\g\d\m\n}=U(p,Q,R,p')_{\g\d\a\b\n\m}.
\end{align}
In effect, one finds
\aeq\label{amplitudestar}
-i\,\M^*=e^2[{\ef}^{\t\x}({\mathbf p},\lambda
)]^*U(p,Q,R,p')_{\t\x\s\r\h\z}\ef^{\s\r}({\mathbf p}',
 \lambda^\prime)\e^{\z}({\mathbf q}^\prime, \ell^\prime )[\e^{\h}({\mathbf
q},\ell)]^*,
\ceq
which allows us to calculate the squared amplitude as
\begin{eqnarray}
\vert \M\vert^2=\M\, \M^*&=&e^4[{\ef}^{\a\b}({\mathbf p}',
\l')]^*U_{\a\b\g\d\m\n}(p',Q,R,p)
\ef^{\g\d}({\mathbf p},\lambda)[{\ef}^{\t\x}({\mathbf p},\lambda )]^*
U(p,Q,R,p')_{\t\x\s\r\h\z}\ef^{\s\r}(p')\nonumber\\
&&\times
\e^{\n}({\mathbf q},\ell)[\e^{\h}({\mathbf q},\ell)]^*
\e^{\z}({\mathbf q}^\prime, \ell^\prime)[\e^{\m}({\mathbf q}^\prime,
\ell^\prime)]^*
\end{eqnarray}
Finally, the averaged squared amplitude is found upon summing over the polarizations (abbreviated as $pol$) and amounts to
\aeq\label{m2}
\overline{\vert\M\vert^2}=\frac{1}{6}\sum_{pol}\vert\M\vert^2=\frac{e^4}{6}{\mathbf
P}^{\s\r\a\b}({\mathbf p}')
U_{\a\b\g\d\m\n}(p',Q,R,p){\mathbf P}^{\g\d\t\x}({\mathbf
p})U(p,Q,R,p')_{\t\x\s\r}{}^{\n\m}.
\ceq
Here,  use was made of
\aeq\label{fproj}
\sum_\ell{\e^{\n}({\mathbf q},\ell)[\e^{\h}({\mathbf q},\ell)}]^*=-g^{\n\h},\qquad
\sum_{\ell' }{\e^{\z}({\mathbf q}', \ell' )[\e^{\m}({\mathbf
q}',\ell')}]^*=-g^{\z\m},
\ceq
and
\aeq
\sum_{\l} \ef^{\g\d}({\mathbf p},\l)[{\ef}^{\t\x}({\mathbf p},\l)]^*
={\mathbf P}^{\g\d\t\x}({\mathbf p}),\qquad
\sum_{\l'}\ef^{\s\r}({\mathbf p}' ,\l')[{\ef}^{\a\b}({\mathbf p}' ,\l')]^*
={\mathbf P}^{\s\r\a\b}({\mathbf p}^\prime ).
\ceq
An essential ingredient of the amplitude is the ${\mathbf P}^{\g\d\t\x}(p)$ projector  operator which identifies the properties of the  degrees of freedom in $(1,0)\oplus (0,1)$ on which the scattering of photons takes place.  The latter, here denoted by $w_+^{\a\b}({\mathbf p}, \l)$ and $w_-^{\a\b}({\mathbf p},\l)$, are in their turn the eigenstates to the parity operator and of  either positive or negative eigenvalues.

\subsection{\label{sec3b}Compton scattering off the parity degrees of freedom}
\subsubsection{Poincar\'e covariant projector method}
Before commencing we recall the projectors on the states of positive \{$(w_+^{\alpha\beta}(\mathbf{p} ,\l)$\}, and negative [$w_-^{\alpha\beta }(\mathbf{p} ,\l) $] states from the respective Eqs.~(\ref{par_T_plus}) and (\ref{par_T_minus}), 
\begin{eqnarray}
{\mathbf P}^{\a\b\d\g}_+(\mathbf{p})=
\sum_{\l}w_+^{\a\b}(\mathbf{p},\l)[{w}_+^{\d\g}(\mathbf{p},\l)]^*, &\quad&
{\mathbf P}^{\a\b\d\g}_-(\mathbf{p})=
-\sum_{\l}w_-^{\a\b}(\mathbf{p},\l)[{w}_-^{\d\g}(\mathbf{p},\l)]^*.
\label{prjtrs}
\end{eqnarray}

The Compton scattering amplitudes off the parity degrees of freedom are then obtained upon subsequently substituting ${\mathbf P}^{\a\b\d\g} ({\mathbf p})$, and ${\mathbf P}^{\a\b\d\g} ({\mathbf p}^\prime )$ in (\ref{m2}) by one of the projectors in (\ref{prjtrs}) according to
\begin{eqnarray}
\label{m2r1}
\overline{\vert\M\vert^2}&=&\sum_{\pi_1}M_{\pi_1\pi_1},\\
M_{\pi_1\pi_1}&=&\frac{e^4}{6}{\mathbf P}^{\s\r\a\b}_{\pi_1}({\mathbf p}' )
U_{\a\b\g\d\m\n}(p',Q,R,p){\mathbf P}^{\g\d\t\x}_{\pi_1}({\mathbf p})
 U(p,Q,R,p')_{\t\x\s\r}{}^{\n\m}, \quad \pi_1=+, -,
\end{eqnarray}
Specifically, for $\pi_1=-$, the following expression is  obtained:
\aeq\label{mmm}
M_{--}=f_0+\frac{8}{3}f_D+f_T
+\frac{e^4(2 m^2-s-u)}{12 m^4 \left(m^2-s\right)^2
\left(m^2-u\right)^2}\sum_{l=1}^4 ({\widetilde g}-2)^l a^{--}_{l}.
\ceq
Here, we have introduced the notations,
\begin{eqnarray}
f_0&=&\frac{4e^4(5m^8-4(s+u)m^6+(s^2+u^2)m^4 +s^2u^2)}{(m^2-s)^2(m^2-u)^2},\label{f0}\\
f_D&=&-\frac{2e^4(-2m^2 +s+u)^2}{(m^2-s)(m^2-u)},\label{fd}\\
f_{T}&=&\frac{2e^4 \left(-2 m^2+s+u\right)^2}{m^4}\label{fva},
\end{eqnarray}
with $s$, $t$, and $u$ being the standard Mandelstam variables.

It is interesting to notice that
\begin{itemize}
\item the terms $f_0$ and $f_D$ are finite in the ultrarelativistic limit,
\item the $f_T$ term diverges at high energies,

\item all the ${\widetilde g}$ dependence enters the expression in  the form of a power series of $({\widetilde g}-2)^l$, the expansion coefficients $a_{l}^{--}$ being calculated as
\begin{align}\label{fDs}
a^{--}_1=&\,-16 \left(m^2-s\right) \left(m^2-u\right) \left(2 m^2-s-u\right)
\left(3 m^4+3 (s+u) m^2-5 s u\right),\\
a^{--}_2=&\,4 \left(-30 m^{10}+18 (s+u) m^8+\left(19 (s^2+u^2)+66 u s\right)
 m^6\right.\nonumber\\
&\,\left.-(s+u) \left(11 (s^2+u^2)+100 u s\right) m^4+7 s u \left(5
(s^2+u^2)+16 u s\right)
m^2-24 s^2 u^2 (s+u)\right),\\
a^{--}_3=&\,-16 \left(m^2-s\right) \left(m^2-u\right) \left(3 m^6-\left(s^2+7
u s+u^2\right)
m^2+3 s u (s+u)\right),\\
a^{--}_4=&\,-6 m^{10}+9 (s+u) m^8-2 (s+u) \left(s^2+10 u s+u^2\right) m^4\\
&+2 s u \left(5 s^2+17 u s+5 u^2\right) m^2-9 s^2 u^2 (s+u).
\end{align}
\end{itemize}
With that in mind, the immediate conclusions following from  (\ref{mmm}) are that
\begin{itemize}
\item the divergence brought about by the ${\widetilde g}$ dependent terms can
be removed by setting
${\widetilde g}=2$,
\item the divergence due to the parameter-independent term $f_T$ is unavoidable.
\end{itemize}
For $M_{++}$ same expression is found, i.e.
\begin{equation}
M_{++}= M_{--}.
\label{pp_mm}
\end{equation}
\subsubsection{The differential cross section}

The differential cross section is now obtained in terms of  the sum of $M_{--}$, and $M_{++}$ as
\aeq\label{dsd}
\frac{d
\s}{d\Omega}=\(\frac{1}{8\pi m}\frac{\o'}{\o}\)^2[M_{--}+M_{++}]=\(\frac{d\s_{--}}{d\Omega}+\frac{d\s_{++}}{d\Omega}\),
\ceq
where we defined
\aeq
\frac{d\s_{--}}{d\Omega}=\(\frac{1}{8\pi m}\frac{\o'}{\o}\)^2 M_{--},\qquad
\frac{d\s_{++}}{d\Omega}=\(\frac{1}{8\pi m}\frac{\o'}{\o}\)^2 M_{++}.
\ceq
We perform all simplifications in the laboratory  frame, here
\aeq\label{manlab}
s=m(m+2\o),\qquad u=m(m-2\o'),
\ceq
with $\o$ and $\o'$ being the energies of the incident, and the scattered photons, respectively. The $\o$ and $\o'$ variables  are related by means of the scattering angle $\q$ as
\aeq\label{omegas}
\o'=\frac{m\o}{m+\o(1-\cos \q)},
\ceq
after some algebraic manipulations, one can cast the differential cross section derived  from  (\ref{mmm}) as
\aeq\label{dsmm}
\frac{d\s_{--}}{d\Omega}=z_0+\frac{8}{3}z_D+z_T+\frac{(x-1) \eta ^2
r_0^2}{48 (-x \eta +\eta +1)^4}
\sum_{l=1}^{4}{ ({\widetilde g}-2)^l b_l^{--}}.
\ceq
Here, $r_0=e^2/(4\pi m)=\a/m$, $x=\cos\q$ and $\h=\o/m$, while the $z_i$ terms are related to the $f_i$ as $\(\frac{1}{8\pi m}\frac{\o'}{\o}\)^2 f_i$; explicitly we get
\begin{eqnarray}
z_0&=&\frac{\left(x^2+1\right) r_0^2}{2 ((x-1) \eta -1)^2},\label{z0}\\
z_D&=&-\frac{(x-1)^2 \eta ^2 r_0^2}{2 ((x-1) \eta -1)^3}, \label{zd}\\
z_T&=&\frac{2 (x-1)^2 \eta ^4 r_0^2}{(\eta(x-1) -1)^4}\label{zVA}.
\cec
Correspondingly, the coefficients $b_l^{--}$ in the power series expansion of
the ${\widetilde g}$ dependent piece of the cross section relate to $a_i^{--}$ as $b_i^{--}=\frac{1}{8m^6 \o'^3\o}a_i^{--}$, this is
\begin{align}
b^{--}_1=&64 (x-1) \left((x+4) \eta ^2-x \eta +\eta +1\right),\\
b^{--}_2=&8 \left(-x^2+3 x+(x-1) (13 x+35) \eta ^2+(x-1) ((x-3) x+  \eta
-8\right),\\
b^{--}_3=&32 ((x-1) \eta  (2 (x+2) \eta +1)-1),\\
b^{--}_4=&(x-1) (x (x+14)+21) \eta ^2+2 (x-1) (x+3) \eta -2 (x+3).
\label{exp_cff}
\end{align}

Equation (\ref{dsmm}), in combination with the subsequent Eqs.~(\ref{zVA})-(\ref{exp_cff}) shows that the differential cross section is a well-behaved quantity in the forward direction, $x\to 1$, for any ${\widetilde g} $, and divergent elsewhere. This contrasts the situation of Compton scattering off the four-vector, earlier
reported  by us in \cite{Napsuciale:2007ry}, which has been found to be well behaved in all directions provided the gyromagnetic ratio was to take the universal value of $g_V=2$. For spin $1$ in the antisymmetric tensor, the ${\widetilde g}=2$ value solely kills the last term on the right-hand side in (\ref{dsmm}) but one is nonetheless left with the ${\widetilde g}$-independent divergent term $z_T$ in (\ref{zVA}). Therefore, no constraint could be obtained on the ${\widetilde g}$ value which so far still remains undetermined. In effect, unitarity is violated in  Compton scattering off $(1,0)\oplus (0,1)$, and in contrast to same process regarding  the four-vector. The problem extends to the total cross sections as well. In effect, one is forced to admit that massive fundamental spin-$1$ bosons  residing in the antisymmetric tensor present themselves as profoundly distinct from those residing in the four-vector.
\subsubsection{Proca's method}
The differential cross section for Compton scattering from the negative-parity states  in $\tensorrep$ can also be obtained within Proca's method along the line of Sec. \ref{nproca}.  This is done by  replacing the propagator in \eqref{tprop} by the one in \eqref{pmprop}, and the $\Gamma_{\a\b\g\d\m\b}$ tensors in \eqref{Tmntensor} by the $T^{-}_{\a\b\g\d\m\n}$ tensors in \eqref{tmprocat}. In effect, the following result is found:
\aeq\label{dspnp}
\frac{d\s^P_{--}}{d\Omega}=\frac{r_0^2}{96 (-x \eta +\eta +1)^6}
\sum_{l=0}^{6}{\h^l b_l^P}.
\ceq
Here,
\begin{align}
b^{P}_0=&48 \left(x^2+1\right),\\
b^{P}_1=&-192 (x-1) \left(x^2+1\right),\\
b^{P}_2=&2 \left(144 x^4-288 x^3+295 x^2-336 x+187\right),\\
b^{P}_3=&-6 (x-1) \left(32 x^4-64 x^3+71 x^2-112 x+75\right),\\
b^{P}_4=&(x-1)^2 \left(48 x^4-96 x^3+139 x^2-400 x+335\right),\\
b^{P}_5=&-16 (x-1)^3 \left(x^2-8 x+9\right),\\
b^{P}_6=&(x-1)^4 \left(x^2-16 x+29\right).
\end{align}
Analyzing the above expression, one finds the correct low energy classical limit.
However, at high energy
this differential cross section grows infinitely in
all directions. This behavior has to be attributed
to the differences in the vertices defining the Feynman rules  in the two
methods under discussion.

\subsection{\label{cw}Compton scattering off the parity doublet as a whole}
In the current section we shall entertain options for restoring unitarity in Compton scattering off $(1,0)\oplus (0,1)$. We begin with the observation that in substituting ${\mathbf P}^{\a\b\d\g} ({\mathbf p})$ and ${\mathbf P}^{\a\b\d\g} ({\mathbf p}^\prime )$ in (\ref{m2}) by  the projectors on the whole space, describing the spin-$1$ parity doublet according to
\begin{eqnarray}
{\mathbf P}^{\a\b\d\g} ({\mathbf p})={\mathbf P}^{\a\b\d\g}_+(\mathbf{p}) &+&
{\mathbf P}^{\a\b\d\g}_-(\mathbf{p})=\mathbf{1}^{\a\b\g\d},
\label{prjtrs_whole}
\end{eqnarray}
would  generate the following averaged squared amplitude:
\begin{eqnarray}
\label{m2r2}
\overline{\vert\mathcal{M}\vert^2}&=&\sum_{\pi_1,\pi_2}M_{\pi_1\pi_2},\\
M_{\pi_1\pi_2}&=&\frac{e^4}{6}{\mathbf P}^{\s\r\a\b}_{\pi_2}({\mathbf p}' )
U_{\a\b\g\d\m\n}(p',Q,R,p){\mathbf P}^{\g\d\t\x}_{\pi_1}({\mathbf p})
 U(p,Q,R,p')_{\t\x\s\r}{}^{\n\m}, \quad \pi_1=\pm, \quad \pi_2=\pm.
\end{eqnarray}
Because of $M_{++}=M_{--}$ and $M_{+-}=M_{-+}$,  the following expression is obtained:
\begin{equation}
\overline{\vert\mathcal{M}\vert^2}={2}(M_{--}+M_{+-}),
\label{mixed_m}
\end{equation}
where
\begin{eqnarray}
M_{+-}=-\frac{1}{6}\sum_{\textup{pol}}
\left\vert e^2\[w_-^{\alpha\beta}({\mathbf p}^\prime,\lambda^\prime)\]^\ast
U_{\a\b\g\d\m\n}(p,Q,R,p^\prime) w_+^{\g\d}({\mathbf p},\l )
\[\epsilon^\mu (q^\prime,\lambda^\prime)\]^\ast\epsilon^\nu (q,\l)\right\vert^2
\label{mixed_ampl}
\end{eqnarray}
The explicit expression for $M_{+-}$ is now calculated  as
\begin{align}\label{mmp}
M_{+-}=-M_{--}+f_0+\frac{8}{3}f_D+\frac{2({\widetilde g}^2-4)}{3}f_D, \quad
M_{+-}+M_{--}=f_0+\frac{8}{3}f_D+\frac{2({\widetilde g}^2-4)}{3}f_D.
\end{align}
This expression shows that due to the presence of $M_{--}$ in the expression for $M_{+-}$,  the parity-mixed amplitude is also divergent. However, due to the negative sign of $M_{--}$ in (\ref{mmp}), the net amplitude $M_{--}+M_{+-}$ for the parity doublet will be well behaved. Indeed, insertion of the last equation in (\ref{mixed_m}), and  accounting for the fact that  the $f_0$ and $f_D$ quantities in (\ref{fva}) are finite in the ultraviolet, amounts to the desired cancellation of the divergent terms.
\begin{quote}
This is the key result of the present work and a comment on the details staying behind is in order. Notice that the cancellation between the divergences in the parity-diagonal, and the parity-mixed amplitudes, $M_{--}$ and $M_{-+}$ in (\ref{mmp}),  was achieved  on the cost of assuming equality between the  vertices including the photon and the $w_+^{\a\b}\to w_+^{\a\b}$ vector current, on the one side, and the photon and the axial $w_+^{\a\b}\to w_- ^{\a\b}$ (equivalently, $w_+^{\a\b}\to \c w_+ ^{\a\b}$) current on the other side. Here, $\c$ stands for the chirality operator in $(1,0)\oplus(0,1)$ defined in \eqref{chi}. Such is justified  only if the massless projectile that couples in one of the vertices to the parity-mixed  target current is an axial vector, such as an axial photon (see Fig.~\ref{Figure_axialphoton}).
\begin{figure}
\begin{minipage}{0.3\textwidth}
\bigskip
\begin{center}
\includegraphics{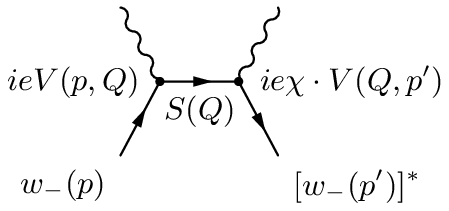}
\end{center}
\end{minipage}
\begin{minipage}{0.3\textwidth}
\bigskip
\begin{center}
\includegraphics{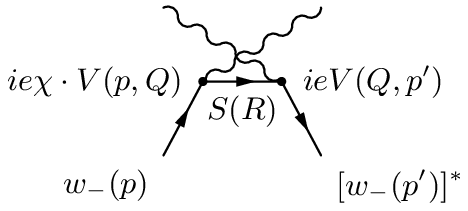}
\end{center}
\end{minipage}
\begin{minipage}{0.3\textwidth}
\bigskip
\begin{center}
\includegraphics{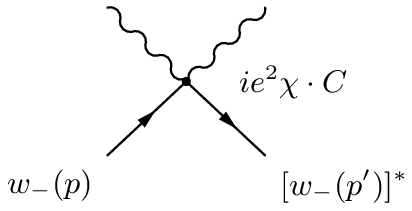}
\end{center}
\end{minipage}
\caption{\label{Figure_axialphoton} Diagrams representing the emission and absorption of a massless axial vector projectile, denoted by $\tilde{\e}(q')$.}
\end{figure}
In effect, the divergence cancellation presented above can take place in one of the possibilities in theories of electromagnetic interactions which allow for two photons of opposite parities, as considered in dual theories of electromagnetism \cite{Mignaco:2001pi,Olive:2000zv}. Alternatively, within Standard Model physics, the place of the axial spin-$1$ projectile could be taken by a gradiently coupled massless Goldstone pion, in which case the process under investigation would correspond to  photo-pion production off a chiral system. The expressions for the cross sections have to be slightly modified algebraically to include the new coupling. The axial current by itself, which underlies the parity-mixed amplitude, $M_{+-}$,  is allowed because by virtue of the commutativity of the parity- and charge-conjugation operators in the representation space under investigation, the opposite parity states do not behave as particles and antiparticles to each other.

\end{quote}

------------------------------------
\subsubsection{The differential cross sections}
The differential cross section in the laboratory  frame follows from
\aeq\label{dsformula}
\frac{d\s}{d\Omega}=\(\frac{1}{8\pi m}\frac{\o'}{\o}\)^2 \overline{\vert\mathcal{M}\vert^2}
=2\(\frac{d\s_{--}}{d\Omega}+\frac{d\s_{+-}}{d\Omega}\),
\ceq
where have defined
\aeq
\frac{d\s_{--}}{d\Omega}=\(\frac{1}{8\pi m}\frac{\o'}{\o}\)^2 M_{--},\qquad
\frac{d\s_{+-}}{d\Omega}=\(\frac{1}{8\pi m}\frac{\o'}{\o}\)^2 M_{+-}.
\label{dsigma_pm}
\ceq

The substitution of (\ref{omegas}) in (\ref{dsformula}) amounts to the following expression for $d \s_{+-}/d\Omega$:
\aeq\label{dspm}
\frac{d\s_{+-}}{d\Omega}=-\frac{d\s_{--}}{d\Omega}+z_0+\frac{8}{3}z_D+\frac{2}{3}({\widetilde
g}-2)({\widetilde g}+2)z_D,
\ceq
with $z_0$ and $z_D$ from \eqref{z0} and \eqref{zd}. Notice that differently from the scattering off the parity states, here

\begin{itemize}
\item all the terms entering the expression for the cross section are finite,
\item the ${\widetilde g}$ parameter looses its importance as a tool for removing divergences.
\end{itemize}

Upon  substitution of (\ref{dsigma_pm}) in (\ref{dsd}),  the final result for scattering off the parity doublet as a a whole (\ref{prjtrs_whole}) is obtained. As we can see  from (\ref{dsmm}) and (\ref{dspm}), both $d\s_{--}/d\Omega $ and $d\s_{+-}/d\Omega$ have a divergent high energy behavior contained both in  the  $z_T$-term in (\ref{zVA}) and the coefficients $b_l^{--}$ in (\ref{exp_cff}). However, because the $d\s/d\Omega $ is defined by the difference of these two divergences, one ends up with a
well-behaved differential cross section in the ultraviolet, as illustrated by Figs. \ref{dsdecomlow} and \ref{dsdecomhigh}. There we plot the quantities
\aeq\label{dss}
d\tilde{\s}\equiv\frac{1}{2 r_0^2}\frac{d\s}{d\Omega},\qquad
d\tilde{\s}_{--}\equiv\frac{1}{r_0^2}\frac{d\s_{--}}{d\Omega},\qquad
d\tilde{\s}_{+-}\equiv\frac{1}{r_0^2}\frac{d\s_{+-}}{d\Omega},
\ceq
again assuming   ${\widetilde g}=2$ for concreteness.
\begin{figure}[ht]
\includegraphics{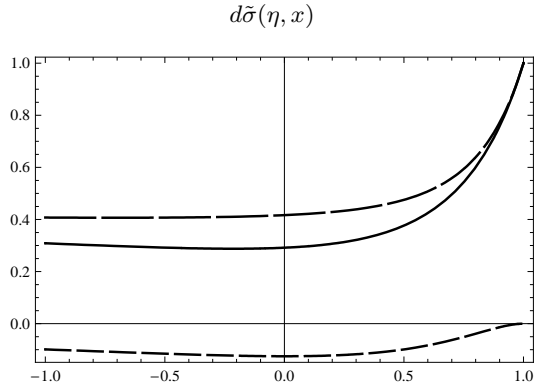}
\caption{\label{dsdecomlow} Plot of $d\tilde{\s}_{--}$ (long-dashed line) and $d\tilde{\s}_{+-}$ (short-dashed line) from Eq. \eqref{dss} together with $d\tilde{\s}=d\tilde{\s}_{--}+d\tilde{\s}_{+-}$ (continuous line), for ${\widetilde g}=2$, at energy $\h=1$, as functions of $x=\cos{\q}$,
where $\q$ is the scattering angle in the laboratory  frame.}
\end{figure}
\begin{figure}[ht]
\includegraphics{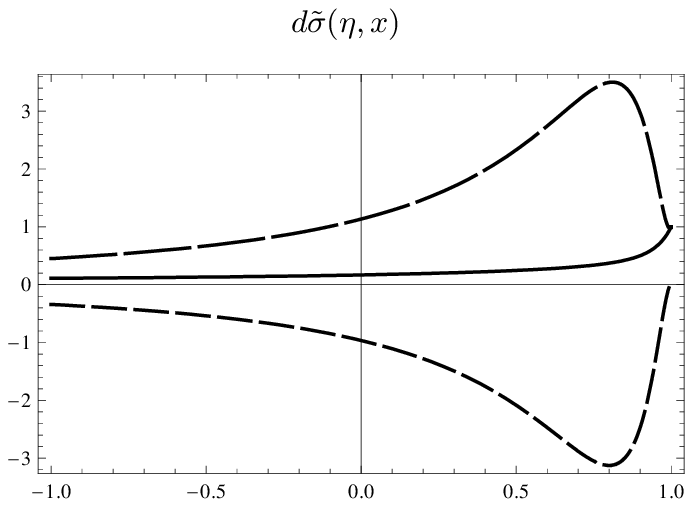}
\caption{\label{dsdecomhigh} Plot of $d\tilde{\s}_{--}$ (long-dashed line) and $d\tilde{\s}_{+-}$ (short-dashed line) from Eq. \eqref{dss} together with $d\tilde{\s}=d\tilde{\s}_{--}+d\tilde{\s}_{+-}$ (continuous line), for ${\widetilde g}=2$, at energy $\h=5$, as functions of $x=\cos{\q}$,
where $\q$ is the scattering angle in the laboratory  frame.}
\end{figure}
\subsubsection{The total cross section}
According to Eq.~(\ref{dsd}), the total cross section accordingly separates as
\aeq
\s=\int_{0}^{2\pi} {d\f\int_{-1}^{1}{d x \frac{d\s}{d\Omega}}}=2(\s_{--}+\s_{+-}).
\ceq
Carrying out the integration amounts to
\aec
\s_{--}&=&s_0+\frac{8}{3}s_D+s_T
+\sum_{l=1}^4 ({\widetilde g}-2)^l\(\frac{A_l^T}{48 \eta  (2 \eta +1)^3}
+\frac{\log (2 \eta +1)B_l^T}{32 \eta ^2}\)\s_T,\label{tot_CSDiag}\\
\s_{+-}&=&-\s_{--}+s_0+\frac{8}{3}s_D+\frac{({\widetilde g}-2)({\widetilde
g}+2)}{3}s_D,
\label{tot_CSmixed}
\end{eqnarray}
where $\s_T$ is the Thompson cross section $\s_T=(8/3)\pi r_0^2$, and
\aec
s_0&=&\frac{3 (\eta +1) \s_T (2 \eta  (\eta +1)
-(2 \eta +1) \log (2 \eta +1))}{4 \eta ^3 (2 \eta +1)},\label{s0}\\
s_D&=&\frac{3 \s_T \left((2 \eta +1)^2 \log (2 \eta +1)
-2 \eta  (3 \eta +1)\right)}{8 \eta  (2 \eta +1)^2},\label{sd}\\
s_T&=&\frac{4 \eta ^4 \s_T}{(2 \eta +1)^3}.
\cec
Furthermore, the explicit evaluation of the  coefficients, $A_l^T$, in the power series expansion of the ${\widetilde g}$ dependent piece of the cross section, i.e. in the last term on the right hand side in (\ref{tot_CSDiag}) gives
\aec
A_1^T&=&32 \eta  \left(20 \eta ^4+22 \eta ^3-3 \eta ^2-12 \eta -3\right),\\
A_2^T&=&4 \left(192 \eta ^5+322 \eta ^4+219 \eta ^3+90 \eta ^2+42 \eta
+9\right),\\
A_3^T&=&16 \eta ^2 \left(24 \eta ^3+50 \eta ^2+33 \eta +6\right),\\
A_4^T&=&\eta  \left(72 \eta ^4+212 \eta ^3+194 \eta ^2+69 \eta +9\right),\\
B_1^T&=&-32 (\eta -1) \eta,\\
B_2^T&=&4 \left(-13 \eta ^2+\eta -3\right),\\
B_3^T&=&-32 \eta ^2,\\
B_4^T&=&-\eta  (8 \eta +3).
\cec
Now in taking the sum of Eqs.~(\ref{tot_CSDiag}) and (\ref{tot_CSmixed}), one sees that the divergent  $\sigma_{--}$ is completely  canceled  by $\sigma_{+-}$ because of the negative  sign of  $\sigma_{--}$ contained in it. The cancellation of the divergent behavior is shown in Figs. \ref{sigmadecomlow} and \ref{sigmadecomhigh}, where the following quantities have been displayed:
\aeq\label{ss}
\tilde \s\equiv \frac{1}{2\s_T}\s,\qquad
\tilde{\s}_{--}\equiv \frac{1}{\s_T}\s_{--},\qquad
\tilde{\s}_{--}\equiv \frac{1}{\s_T}\s_{+-},
\ceq
again choosing  ${\widetilde g}=2$ for concreteness.
\begin{figure}[ht]
\includegraphics{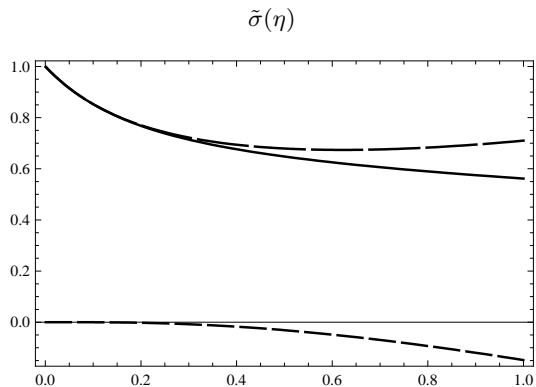}%
\caption{\label{sigmadecomlow}  Plot of $\tilde{\s}_{--}$ (long-dashed line) and $\tilde{\s}_{+-}$ (short-dashed line) from Eq. \eqref{ss} together with $\tilde{\s}=\tilde{\s}_{--}+\tilde{\s}_{+-}$ (continuous line), for ${\widetilde g}=2$ as functions of $\h=\o/m$, where $\omega$ is the energy of the incident photon and $m$ is the mass of the target up to the energy of $\h=1$.}
\end{figure}
\begin{figure}[ht]
\includegraphics{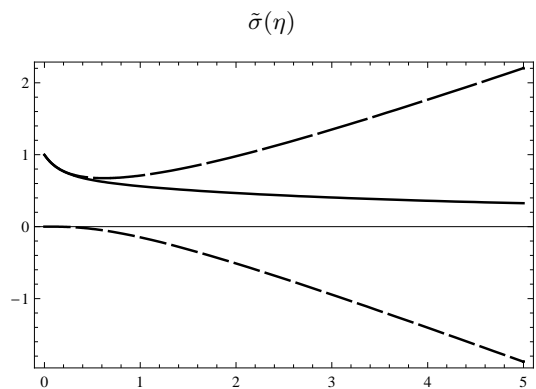}
\caption{\label{sigmadecomhigh} Plot of $\tilde{\s}_{--}$ (long-dashed line) and $\tilde{\s}_{+-}$ (short-dashed line) from Eq. \eqref{ss} together with $\tilde{\s}=\tilde{\s}_{--}+\tilde{\s}_{+-}$ (continuous line), for ${\widetilde g}=2$ as functions of $\h=\o/m$, where $\omega$ is the energy of the incident photon and $m$ is the mass of the target up to the energy of $\h=5$.}
\end{figure}
The extension of the calculation of the Compton scattering cross sections to include the negative-parity degrees of freedom of $(1,0)\oplus (0,1)$ became possible due to their causal propagating character within the framework of the Poincar\'e covariant projector formalism.

\section{Conclusions}

In this work we studied the spin-$1$ description by either the four-vector or the antisymmetric-tensor representation of the Lorentz group, with the special emphasis on the unitarity of  Compton scattering in the ultrarelativistic limit. We observed that the single-parity Proca theory completely fails in describing the process under investigation, leading to cross sections in the ultrarelativistic limit that are divergent in all directions. This shortcoming could be removed within the Poincar\'e covariant projector method, suited equally well for the description of single-parity degrees of freedom on the one side, and of parity doublets on the other. Specifically, the Compton scattering cross section off the parity degrees of freedom in $(1,0)\oplus (0,1)$ turns out to be finite in the forward direction, though it is still divergent elsewhere. Moreover, they turned out to be independent of the value of the gyromagnetic ratio. This behavior is quite different from that of Compton scattering  off spin-$1$ states  transforming as four-vectors, where, within the framework employed here, the cross sections are finite in all directions in the ultraviolet for the universal value $g=2$ of the gyromagnetic ratio.

The price for achieving finite Compton scattering amplitudes for spin $1$ residing in the antisymmetric tensor has been to admit for vertices describing the coupling of a massless axial-vector projectile to the parity-mixed current in the representation under consideration, in which case the net parity-doublet amplitudes have been obtained finite in all directions unconditionally and for any $g$. In one of the possibilities, massless spin-$1$ projectiles could be axial photons, allowed to coexist  with vector photons in dual theories of electrodynamics. Alternatively, their place can be taken by  gradiently coupled Goldstone pions, in which case the expressions for the cross sections  have to be slightly modified algebraically to include the new coupling. Moreover, fundamental spin-$1$ antisymmetric-tensor bosons can also exist as composite hadrons, in which case the aforementioned  divergences could be removed by properly behaving form factors.

We furthermore analyzed in detail the vector-tensor equivalence theorem and showed that it has its validity solely for the unphysical unit value of the gyromagnetic ratio, and  is otherwise violated through the gauging procedure at the level of the genuine hyperbolic and causal gauged equations on the one side, and at the level of  the electromagnetic currents on the other. As long as the electromagnetic current in the four-vector representation is characterized by an oblate electric quadrupole moment, as is known for the $W$ boson,  while the antisymmetric tensor is characterized by a prolate  one,  we conclude that if the  physical $\rho$ meson were to belong to  $(1,0)\oplus (0,1)$, its negative magnetic quadrupole moment has to be attributed to the effects of its internal structure. This is because recent Lattice QCD calculations and associated  measurements \cite{Hedditch:2007ex}, \cite{Dbeyssi:2011ep} strongly speak in favor of an oblate electric quadrupole moment of the $\rho$ meson, and in contrast to previous ones \cite{Cardarelli:1994yq}, \cite{Jaus:2002sv}. In order for a particle residing in $(1,0)\oplus (0,1)$ to carry a negative electric quadrupole moment, its gyromagnetic ratio has to be less than $1$, if such happens to be the effect of the surrounding meson cloud. For the time being, the assignment of the $\rho$ meson to a Lorentz group representation remains an open question. As a further intriguing meson pair,  we wish to mention the $f_1(1420)$-$\omega (1420)$ couple which also may appear as a good candidate for a $(1,0)\oplus(0,1)$ resident. Our result shows that the description of the $(1,0)\oplus (0,1)$ representation space provided by the Poincar\'e covariant projector method is that of a physically observable chiral doublet. We expect this finding to extend to all bosonic pure spin representations, which are most likely to behave as parity doublets. However, the  fermion ones are bound to behave as parity states. The reason is that there the  parities are separated by the fermion quantum number and characterize the respective particles and antiparticles. As a test we  reproduced the calculation  of Ref.~\cite{DelgadoAcosta:2010nx} regarding Compton scattering off a $(1/2,0)\oplus (0,1/2)$ target, but performed it this time  along the lines of Sec. \ref{sec3b}. We observed that the contributions of the corresponding parity-mixed amplitudes are nullified by the universal $g=2$ value of the gyromagnetic ratio,
thus revealing its fundamental importance as a key tool for controlling the fermion number conservation.



\end{document}